\documentclass[epj]{svjourmod}
\usepackage{amsmath,amssymb}
\usepackage{graphicx}

\allowdisplaybreaks[1]

\begin{document}

\newcommand{\sh}{\not\!}
\newcommand{\ol}{\overline}
\newcommand{\rR}{R}
\newcommand{\rr}{{\cal R}}
\newcommand{\mc}{\multicolumn{2}{c}}
\newcommand{\mcl}{\multicolumn{2}{l}}
\newcommand{\mcc}{\multicolumn{3}{c}}
\newcommand{\nl}{\!\!\! & \!\!\!}
\newcommand{\av}[1]{\langle {#1}\rangle}
\def\beq{\begin{equation}}
\def\eeq{\end{equation}}
\def\bea{\begin{eqnarray}}
\def\eea{\end{eqnarray}}
\def\kl3g{K_{l3\gamma}}
\newcommand{\mpi}{M_\pi^2}
\newcommand{\mk}{M_K^2}
\newcommand{\me}{M_\eta^2}
\newcommand{\eq}{\,=\,}
\newcommand{\no}{\nonumber}
\newcommand{\Order}{{\cal O}}
\numberwithin{equation}{section}
\newcommand{\scs}{\, , \,}
\newcommand{\sem}{\, ; \,}
\newcommand{\nnnl}{\nonumber\\}
\newcommand{\fs}{\, . \,}
\newcommand{\vibf}{\rule[3.8mm]{3mm}{.1mm} \hspace{-3mm}V^{\,\rm IB}_{\mu\nu}}
\newcommand{\vsdf}{\rule[3.8mm]{3mm}{.1mm} \hspace{-3mm}V^{\,\rm SD}_{\mu\nu}}
\newcommand{\vmn}{V_{\mu\nu}}
\newcommand{\Eg}{E_\gamma^*}
\newcommand{\Ecut}{E_\gamma^{\rm cut}}
\newcommand{\Ep}{E_\pi^*}
\newcommand{\Ee}{{E_e^*}}
\newcommand{\te}{\theta_{e\gamma}^*}
\newcommand{\tp}{\theta_{\pi\gamma}^*}
\newcommand{\tecut}{\theta_{e\gamma}^{\rm cut}}
\newcommand{\vsdifc}[1]{\widetilde V_{#1}}

\hyphenation{brems-strahlung}
\hyphenation{struc-ture}

\title{Radiative \boldmath{$K_{e3}$} decays revisited}
\author{
J. Gasser\inst{1}, B. Kubis\inst{1,2}, N. Paver\inst{3},
 {\rm and} M. Verbeni\inst{4}\\
}
%
%
\institute{Institut f\"ur theoretische Physik, Universit\"at Bern, 
Sidlerstrasse 5, CH-3012 Bern, Switzerland
\and
 Helmholtz-Institut f\"ur Strahlen- und Kernphysik,
    Universit\"at Bonn, Nussallee 14-16, D-53115 Bonn, Germany
\and
 Dipartimento di Fisica Teorica,
    Universit\`a degli Studi di Trieste,
and INFN-Trieste, Strada Costiera 11, I-34100 Trieste, Italy
\and
 Departamento de F\'\i sica Te\'orica y del Cosmos,
    Universidad de Granada,
 Campus de Fuente Nueva, E-18002 Granada, Spain
}

\date{
}
\abstract{
Motivated by recent experimental results and ongoing measurements, 
we review the chiral perturbation theory prediction for  
$K_L\to\pi^{\mp} e^{\pm} \nu_e\gamma$ decays. Special emphasis is given to the 
stability of the inner bremsstrahlung-dominated relative branching ratio versus 
the $K_{e3}$ form factors, and on the separation of the structure-dependent 
amplitude in differential distributions over the phase space. For the
structure-dependent terms, an assessment of the order $p^6$ 
corrections is given, in particular, 
a full next-to-leading order calculation 
of the axial component is performed. The experimental analysis of the photon 
energy spectrum is discussed, and other potentially useful distributions are 
 introduced.
\PACS{
      {13.20.Eb}{Radiative semileptonic decays of $K$ mesons}
      \and
      {11.30.Rd}{Chiral symmetries}
      \and
      {12.39.Fe}{Chiral Lagrangians}
     } 
} 
\titlerunning{Radiative $K_{e3}$ decays revisited}
\maketitle

\tableofcontents

\section{Introduction} 

The amplitude for the semileptonic radiative decays 
$K_L \!\to\pi^{\mp} l^{\pm} \nu_l\gamma$
$\left[ K_{l3\gamma}\right]$, with $l=e,\mu$, can be divided into two 
components: the inner bremsstrahlung (IB) that accounts for photon radiation 
from the external charged particles and which is determined by the 
non-radiative process $K_L\to\pi^{\mp} l^{\pm} \nu_l$ $\left[K_{l3}\right]$; 
and the structure-dependent (SD) amplitude, also called ``direct emission'', 
that describes photon radiation from intermediate hadronic states and 
represents genuinely new information with respect to the IB one.

Low's theorem~\cite{low}, applied to $K_{l3\gamma}$, states that the 
leading contributions in the expansion of the amplitude in powers 
of the photon four-momentum $q$, namely, the orders $q^{-1}$ and $q^0$, are 
completely determined in a model-independent way by the IB via the 
$K_{l3}$ form factors and their first order derivatives. The SD amplitude 
is then defined by the terms of order $q$ and higher. In 
\cite{FFSprl,FFS}, 
the procedure of~\cite{low} was followed to derive the $q^{-1}$ and 
$q^0$ terms of the IB amplitudes for $K_{l3\gamma}$; moreover, a qualitative, 
model-dependent, assessment of the SD amplitudes was performed using vector 
meson dominance.
In \cite{doncel}, the radiative decay modes 
for both  charged and neutral kaons were calculated,  
taking into account IB terms only.
Originally, the main interest was the precision test of 
soft-photon theorems, allowed by the dominance of IB and the fact that, for 
$K_{l3\gamma}$, the non-radiative amplitude could in principle be studied 
extensively and with high accuracy.

Later,  $K_{l3\gamma}$ decay amplitudes (including charged kaon ones)
 were calculated at  order $p^4$ in chiral perturbation theory (ChPT)
 in \cite{BEG},  
 and branching ratios were evaluated for $l=e,\mu$ in a feasi\-bili\-ty 
study for DAF\-NE~\cite{DAFNE}. An error analysis  and a dedicated study of decay
distributions was postponed to a later stage when precise data would become
 available. It is one of the aims of the present work to provide 
 such an analysis.

ChPT allows for a systematic 
expansion of transition amplitudes for low momenta of the pseudoscalar 
mesons \cite{wein,GL}.       
The lowest order amplitude is only of the IB-type with constant 
$K_{l3}$ form factors, and is  independent of free parameters. 
In addition to providing momentum dependence of the $K_{l3}$ form factors
in the bremsstrahlung, the ${\cal O}(p^4)$ terms
predict the existence of non-vanishing SD amplitudes (vector and 
axial-vector), unambiguously calculable in terms of loop diagrams, 
low-energy constants of the strong  Lagrangian 
${\cal L}_4$~\cite{GL}, and the chiral anomaly~\cite{abb,wzw}. While the 
anomaly does not require new physical parameters, the low-energy constants 
are numerically already well-determined from other, independent, meson 
pro- \linebreak cesses. 
Consequently, the experimental verification of the SD amplitude 
currently represents a significant test of ChPT and, ultimately, of QCD. 
Of course, since the expansion of the SD amplitudes starts at 
${\cal O}(p^4)$, one may inquire about the role of higher order 
 corrections, a point that will be addressed in the sequel.

In practice, this experimental analysis is complicated by the fact that the 
radiative $K_{l3}$ branching ratio is largely dominated by the IB, while the 
SD contribution via IB--SD interference is expected to be an effect 
at the percent-level (the pure SD rate is negligibly small). 
On the other hand, the characteristic $q^{-1}$ behavior of the IB by far 
dominates the lower (and intermediate) photon energy range~\cite{doncel},
while in the upper range where the SD effects become more significant, the  
number of events is severely reduced. In addition,  
precise knowledge of the $K_{l3}$ form factors is required for a 
reliable fit to the photon energy distribution, in order to improve the 
sensitivity to signals of SD contributions through deviations from the pure IB. All that 
calls for high precision measurements of {\it both} $K_{l3\gamma}$ 
{\it and} $K_{l3}$.

The first measurement of the decay $K_L\rightarrow \pi^\pm e^\mp\nu\gamma$ 
with significant statistics 
 was performed by the NA31 Collaboration~\cite{leber} 
in a pioneering experiment, which proved the possibility of  precision
measurements  of this process. Their result for the decay rate 
relative to $K_{e3}$
 decays and for the branching ratio  agreed
with the theoretical predictions 
of \cite{doncel} and \cite{BEG}, respectively. 
 A few years later, the KTeV Collaboration~\cite{ktev} 
determined  the relative branching ratio, together with the photon spectrum, 
at percent-level 
sensitivity. They found a result  which 
is {``significantly lower than all 
 published theoretical predictions''}. Moreover, from the measured 
photon energy distribution, 
two particular combinations of the SD 
amplitudes were obtained, within rather large
uncertainties. Quite recently, new experimental 
results on $K_{e3\gamma}$ with percent-level accuracy have been 
presented by the NA48~\cite{na48} and by the KTeV~\cite{ktev_04} 
Collaborations, respectively.

The experimental situation has therefore become quite promising and clearly 
justifies renewed interest in radiative $K_{e3}$ decays. In this regard, 
particularly relevant topics are the stability of IB with respect to 
the $K_{e3}$ form factor parameterizations, the separation itself of the 
radiative amplitude into IB and SD contributions and the re-visitation
the ChPT calculation 
of the SD component, including in particular next-to-leading 
${\cal O}(p^6)$ corrections. Moreover, this represents an opportunity to 
discuss, besides the photon spectrum, also other differential distributions 
that may help in the study of the SD terms in high statistics 
experiments.

In the sequel, we limit our consideration to the $K_{e3\gamma}$ transition, 
since current experimental data with appropriate statistics refer to this mode 
only. 
Specifically, in Sects.~\ref{sec:exp} and \ref{sec:kin} 
we review the experimental situation and the experimental observables, 
the kinematics, and the amplitude definitions with particular emphasis
on the separation into IB and SD contributions. 
In Sect.~\ref{sec:chpt} we present the ChPT results for the SD amplitudes, 
notably the \linebreak ${\cal O}(p^6)$ corrections for the axial amplitudes.
 In Sect.~\ref{sec:r} we 
numerically discuss the relative radiative branching ratio, while 
Sects.~\ref{sec:sdnum} and \ref{sec:distributions} 
are devoted to numerical estimates of the SD amplitudes, 
the photon energy distribution and the comparison with 
experimental results. Also, other kinds of differential distributions are 
discussed there. 
Finally, Sect.~\ref{sec:conclusions} contains a summary of the results, 
while  details of the calculations are collected in the appendices.

\section{Experimental status and observables\label{sec:exp}}

As already mentioned, we concentrate on  $K_{e3\gamma}$ decays where high 
statistics experimental data on the branching ratio and photon 
energy spectrum have become available. For a presentation of the 
experimental status in the other channels, we refer the reader to the 
PDG listing~\cite{PDG}.

Measuring the decay rate relative to $K_{e3}$  is much safer than  absolute 
measurements, as the former 
is free from uncertainties related to experimental 
normalizations, calibrations, and machine luminosity. Basically, an 
inclusive $K_{e3}$ sample of events is collected, all characterized by 
one lepton and one pion of opposite charges emitted from a 
common vertex, without any restriction on the number of emitted photons. A 
radiative $K_{e3\gamma}$ subsample is extracted by imposing additional 
criteria dictated by the apparatus and the experimental conditions, in 
particular by the request of having at least one hard photon in each of 
those events.

To achieve optimal identification of the candidate $K_{e3\gamma}$ events, 
kinematical cuts are applied to the radiative sample. In
particular,
thresholds in the photon energy and in the photon--electron opening angle are 
usually imposed, see, e.g., \cite{doncel}. Then, the experimental 
results generally concern the relative branching ratio
\begin{equation}
\rR\left(\Ecut, \, \tecut \right) \eq
\frac{
\Gamma\left(K_{e3\gamma}, \, \Eg > \Ecut, \, \te > \tecut \right)
}{
\Gamma\left(K_{e3}\right)  
} ~, \label{def:r}
\end{equation}
where $\Eg$ and $\te$ indicate the photon energy and the 
photon--electron opening angle in the kaon rest frame, respectively.

From the above, the measured value of $\rR$ is determined by the 
ratio of the  number of events in the $K_{e3\gamma}$ and $K_{e3}$ 
 samples, each divided by the respective experimental acceptances. 
The available experimental 
results  are displayed in Table~\ref{tab:experiments}.
\begin{table} 
\centering
\caption{Experimental values of $\rR$ for the transition $K_{e3\gamma}$. The
  first error is statistical, the second one systematic.
  \label{tab:experiments}
}
\medskip
\renewcommand{\arraystretch}{1.5}
\begin{tabular}{ccccc}
\hline 
${\rm Ref.}$ & $\Ecut$ & $\tecut$ & 
events & $\rR \times 10^2 $ \\
\hline
\cite{na48} & $30\hskip 3pt{\rm MeV}$ &  $20^\circ$ & $18977$ & 
$ \,0.964\;\pm 0.008 \; ^{+~0.011}_{-~0.009} ~   $ \\  
\cite{ktev_04}       & $30\,{\rm MeV}$ & $20^\circ$ & $ 4309$ & 
$ 0.916\pm 0.017   $ \\  
\cite{ktev_04} & $10\,{\rm MeV}$ & $   0^\circ   $ & $14221$ & 
$ \,4.942\;\pm 0.042  \pm 0.046$ \\  
\cite{ktev}          & $30\,{\rm MeV}$ & $20^\circ$ & $15463$ & 
$ \,0.908\;\pm 0.008 \;^{+~0.013}_{-~0.012} ~   $ \\  
\cite{leber}            & $30\,{\rm MeV}$ & $20^\circ$ & $ 1384$ & 
$ \,0.934\; \pm 0.036 \; ^{+~0.055}_{-~0.039} ~   $ \\ 
\cite{peach}            & $15\,{\rm MeV}$ & $0^\circ$ &
 $ 10$ & 
$ \,3.3\; \pm 2.0  $ \\ 
\hline
\end{tabular}
\renewcommand{\arraystretch}{1.0}
\end{table}

One may notice that the KTeV 04 result for $\rR$ \cite{ktev_04} with angular 
cuts is based on 
a much smaller number of $K_{e3\gamma}$ events with respect to their previous 
determination \cite{ktev}, yet the achieved uncertainty based on their more recent data 
is comparable owing to a much 
reduced systematic uncertainty.

As regards the photon energy distribution, one can investigate the
spectrum with free normalization because the essential features lie
exclusively in the shape.  
Indeed, structure-dependent emission 
should manifest itself in the harder portion of the photon energy 
spectrum, via a modification of the pure IB spectrum which is 
controlled by the $K_{e3}$ form factors. 
An attempt along these lines was made by the KTeV 01  
experiment~\cite{ktev}, which used a simplified decomposition of the 
structure-dependent amplitude, and their analysis will be commented upon in 
the sequel. This is the only experimental information on structure-dependent 
emission currently available, as neither NA48 nor KTeV have so far 
presented an analysis of this point based on the more recent data.

As far as the perspectives of $K^\pm_{e3\gamma}$ measurements are concerned 
that may complement the currently available data for $K_{e3\gamma}$, new 
results on these transitions are expected from the NA48 
experiment~\cite{NA48private}.
In the more remote future, substantially increased statistics for 
$K^\pm_{l3\gamma}$ decays should be expected in connection with the 
construction of higher intensity kaon beams, with accumulated samples of 
$10^6$ (or more) candidate events~\cite{landsberg}.

\section{The decay amplitude}\label{sec:kin}

In the following, we consider the decay 
\bea
K^0(p)&\to& \pi^-(p')\,e^+(p_e)\,\nu_e(p_\nu)\,\gamma(q) \qquad [K_{e3\gamma}^0]
\eea
and its charge conjugate mode. 
We disregard $CP$-violating contributions,
 and study the emission of a real photon ($q^2=0$).

\subsection{The matrix element}

\begin{figure}
\vskip 2mm
  \centering
  \includegraphics[width=7.5cm]{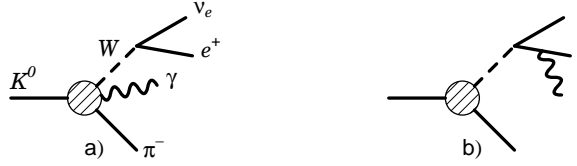}
  \caption{Diagrams describing $\kl3g^0$ decay}
  \label{fig:diag_K0}
\end{figure}

The transition matrix element has the form
\beq\begin{split}
T(K_{e3\gamma}^0) &\eq 
\frac{G_F}{\sqrt{2}}\,e\,V^*_{us}\,\epsilon^\mu(q)^*
\biggl[\bigl(V_{\mu\nu} - A_{\mu\nu}\bigr) \times \\
& \hskip 3.5cm \times
  \bar{u}(p_\nu)\,\gamma^\nu\,(1-\gamma_5)\,v(p_e)\\
&\,+\, \frac{F_\nu}{2 p_e q}\,\bar{u}(p_\nu)\,
\gamma^\nu\,(1-\gamma_5)\,
\bigl(m_e-\sh p_e-\sh q\bigr)\,\gamma_\mu\,v(p_e)\biggr]\\
&\,\doteq\, \epsilon^\mu(q)^* M_\mu ~.
\label{me_Kpg}
\end{split}\eeq
The relevant diagrams are displayed in  Fig.~\ref{fig:diag_K0}.
The first term of \eqref{me_Kpg} corresponds to diagram a), which
  includes bremsstrahlung off the charged pion, 
 while the second one corresponds to the
radiation off the positron, represented by the diagram b). 
 We have introduced the hadronic tensors $V_{\mu\nu}$ and $A_{\mu\nu}$,
\beq \begin{split}
I_{\mu\nu} &\eq i\,\int d^4 x\,e^{iqx}\,
\langle \pi^-(p')|T\,V^{\rm em}_\mu(x)\,I^{\rm had}_\nu(0)|K^0(p)\rangle ~\sem\\
\quad I &\eq V,A  ~,
\end{split} \label{tensor}
\eeq
whereas  $F_\mu$ is the $K_{l3}$ matrix element
\beq
F_\mu\eq \langle \pi^-(p')|V^{\rm had}_\mu(0)|K^0(p)\rangle ~\scs 
\label{me_Kp}   
\eeq
with 
\beq \begin{split}
V_\mu^{\rm had} &\eq \bar{s}\gamma_\mu u ~,~~
A_\mu^{\rm had} \eq  \bar{s}\gamma_\mu \gamma_5 u ~, \\
V^{\rm em}_\mu &\eq \bigl(2\bar u u -\bar d d - \bar s s\bigr)/3 ~.
\end{split}\label{current}
\eeq
The tensors $V_{\mu\nu}$ and $A_{\mu\nu}$ satisfy the 
Ward identities
\beq\begin{split}
q^\mu V_{\mu\nu}&\eq F_\nu ~,\\
q^\mu A_{\mu\nu} &\eq 0 ~,
\label{WI}
\end{split}\eeq
which imply gauge invariance of the total amplitude
\eqref{me_Kpg},
\beq
q^\mu M_\mu\eq 0 ~.
\label{WI_me}
\eeq  

\subsection{Inner bremsstrahlung, structure-dependent\\ 
            terms and all that\label{sec:ibsd}}

Low's theorem is employed
in \cite{FFS} to obtain the IB amplitude for $\kl3g$
decays, written entirely in terms of the $K_{l3}$ form factors and their
derivatives.
In this subsection we present an alternative way of separating the amplitude
into an IB and a SD part  that directly starts from the
following two requirements: 
\begin{enumerate}
\item In order to describe two different physical mechanisms,
the IB and SD amplitudes must be separately gauge invariant.
\item The SD amplitude contains terms of order $q$ and \linebreak higher.
\end{enumerate}
The second condition does not prevent the IB amplitude from
containing terms of order $q$ and higher. Splitting the amplitude under this
less restrictive condition allows one to put more terms into the IB part, still
using only the non-radiative matrix element in this part of the
amplitude. This has the advantage that one can make more precise predictions
for the decay process, as we will see below.

The splitting of the transition amplitude $T$ into an IB and a SD part
requires a corresponding splitting of the hadronic tensor $I_{\mu\nu}$.
Consider first the axial correlator. There are no contributions where the
photon is emitted from the pion line.
Therefore, $A_{\mu\nu}$ is
considered to be a purely SD contribution. 
It can be written in 
the form~\cite{BEG}
\beq \begin{split}
A_{\mu\nu}&\eq A_{\mu\nu}^{\rm SD} \\
&\eq i\,\epsilon_{\mu\nu\rho\sigma}\,
\bigl(A_1\,p'^\rho q^\sigma + A_2\,q^\rho W^\sigma\bigr) \\
&\hskip 0.45cm +i\,\epsilon_{\mu\lambda\rho\sigma}\,
p'^\lambda q^\rho W^\sigma\,\bigl(A_3\,W_\nu+ A_4\,p'_\nu\bigr) ~, 
\end{split} \label{deca}
\eeq
where $W_\mu=(p-p'-q)_\mu$.
(We use the convention $\epsilon_{0123}=+1$.)
The amplitude is manifestly of order $q$ and higher, because the $A_i$ are
non-singular at zero photon energy.
 The Lorentz invariant components  $A_i$ depend on three independent scalar 
variables that can be
 built from $p,\,p'$, and $q$ -- we come back on this in the following paragraph.

The decomposition of the vector correlator reads
\bea\label{eq:decompIBSD}
V_{\mu\nu}=V^{\rm IB}_{\mu\nu}+V^{\rm SD}_{\mu\nu} ~\scs
\eea
where the IB piece is chosen such that 
\bea\label{eq:WIIB}
q^\mu V_{\mu\nu}^{\rm IB}=F_\nu(t)~\scs
\eea
as a result of which we have
\bea\label{eq:WISD}
q^\mu V_{\mu\nu}^{\rm SD}=0~\fs
\eea
The structure-dependent
part of the decay amplitude $T$ in \eqref{me_Kpg} is defined to be
\beq \begin{split}
T^{\rm SD} &\eq \frac{G_F}{\sqrt{2}}\,e\,V^*_{us}\,\epsilon^\mu(q)^*
\left(V_{\mu\nu}^{\rm SD}-A_{\mu\nu}^{\rm SD}\right) \times \\
& \hskip 2.5cm 
\times \bar{u}(p_\nu)\,\gamma^\nu\,(1-\gamma_5)\,v(p_l) ~,
\end{split}
\eeq
whereas the bremsstrahlung part is $T^{\rm IB}=T-T^{\rm SD}$.

It remains to explicitly construct the decomposition
\eqref{eq:decompIBSD}. In order not
to interrupt the argument, we refer the interested reader 
to  Appendix~\ref{app:WI} and simply display here the result,
\begin{align}
V_{\mu\nu}^{\rm IB} &\eq \frac{p'_\mu}{p'q}
\bigl(2p_\nu f_+(W^2)-W_\nu f_2(W^2)\bigr) \no\\[1mm]
&\,+
\frac{W_\mu}{qW}\bigl(2 (p-q)_\nu\triangle f_+-W_\nu\triangle f_2\bigr)
\label{eq:VIB} \\[1mm] &\,
+g_{\mu\nu}\bigl(2\triangle f_+ - f_2(t)\bigr)~\scs \no\\[1mm]
\triangle f_i &\eq f_i(t)-f_i(W^2) ~\scs~ i=+,2 ~\scs
\end{align}
where $f_+,\,f_2$ are the form factors \eqref{me_Kp}
\beq\label{eq:decf}
F_\mu=2p_\mu f_+(t)+(p'-p)_\mu f_2(t) ~\scs~ t=(p-p')^2  ~\fs
\eeq
We  use the form factors $f_+,\,f_2$ instead of the usual
$f_+$, $f_- =f_+-f_2$ ones for easier comparison with the work of~\cite{FFS}.

The IB part derived in \cite{FFS} 
 differs from the one used here through terms of order $q$. It can be obtained
 from $V_{\mu\nu}^{\rm IB}$ by subtracting all terms of order $q$
 and higher from the latter, and merging them into the SD part of the
 amplitude. Because the
 terms to be subtracted can be expressed through the form factors
 $f_+,\,f_2$ and their derivatives, we believe that it does not make much sense
 to perform this purification of the IB part, and we will mostly
 stick with the convention \eqref{eq:VIB}.
  While comparing with the KTeV result~\cite{ktev}, we will have 
 the occasion to compare \eqref{eq:VIB} with the conventional 
 decompositions~\cite{FFS} in more detail in 
 Sect.~\ref{sec:ktevCD}.\footnote{Our separation into IB and SD
 contributions is very
 close in spirit to the notion of \emph{generalized bremsstrahlung} as
 developed in~\cite{genbrems}.}

Let us shortly discuss the salient features of the IB term
\eqref{eq:VIB}. First, it satisfies the Ward identity \eqref{eq:WIIB}. Second, 
it contains all infrared singular pieces
proportional to $1/p'q$. With this we mean the following.
The residue of the singularity is a non-trivial
function of the momenta $p$, $p'$, $q$. The decomposition \eqref{eq:VIB} 
takes into account all singularities at $p'q=0$, in contrast to the
 standard treatment~\cite{FFS}, which considers e.g.\ a term
like  $(qW)^2/p'q$ 
to be of order $q$, to be  relegated to the SD part of
the amplitude. 

It is useful to decompose also  the SD part of the vector amplitude
into a set of gauge invariant tensors.
In the following, we often use the basis  proposed
in~\cite{poblaguev},
\beq \begin{split} 
V_{\mu\nu}^{\rm SD} &\eq
V_1\bigl(p'_\mu q_\nu-p'\!q\,g_{\mu\nu}\bigr)+
V_2\bigl(W_\mu q_\nu -qWg_{\mu\nu}\bigr) \\
&\hskip 0.45cm + V_3\bigl(qWp'_\mu W_\nu-p'\!q\,W_\mu W_\nu\bigr) \\
&\hskip 0.45cm + V_4\bigl(qWp'_\mu p'_\nu-p'\!q\,W_\mu p'_\nu\bigr) ~.
\end{split} \label{eq:ViSD}
\eeq
The Lorentz invariant amplitudes $V_i$ again depend on the 3 scalars that can be
formed from $p$, $p'$, and $q$.

\subsection{Kinematics}

It remains to shortly recall the kinematics of this decay,
 and we begin 
 with the Lorentz invariant amplitudes $A_i,V_i$. As already mentioned, these 
 are functions of three scalar variables that we often take to be
\beq
s \eq (q+p')^2 ~,~~ t\eq (p-p')^2 ~,~~  u\eq (p-q)^2 ~.
\eeq
These variables are useful in the discussion of the analytic properties of $V_i,A_i$.
 In \eqref{tensor}, the variables $s$, $t$, $u$ can assume any value, 
 whereas the  physical region  in $K_{e3\gamma}$ decays
can be represented as follows.
For fixed $W^2$, the variables $s$, $t$, and $u$ vary in
\begin{align}
W^2 & \,\leq\, t \,\leq\, (M_K-M_\pi)^2  ~,\nnnl
s_- & \,\leq\, s \,\leq\, s_+ ~,\nnnl
s_\pm& \eq M_\pi^2-\frac{1}{2t}(t+M_\pi^2-M_K^2)(t-W^2) \nnnl
& \,\pm\, \frac{1}{2t}
\lambda^{1/2}(t,M_K^2,M_\pi^2)\lambda^{1/2}(t,0,W^2) ~,\nnnl
s+t+u&\eq M_K^2+M_\pi^2+W^2 ~,
\label{eq:physicalstu}
\end{align}
where
\beq
\lambda(x,y,z) \eq x^2+y^2+z^2-2(xy+xz+yz) ~.
\eeq
Varying the invariant mass squared $W^2$ 
of the lepton pair in the interval
\beq
m_e^2 \,\leq\, W^2 \,\leq\, (M_K-M_\pi)^2
\eeq
generates the  region covered by  
$s,t,u$ in $\kl3g$ decays. In Sect.~\ref{sec:p6}, 
where the analytic properties
of the amplitudes $A_i,\,V_i$ are discussed, we display 
the region \eqref{eq:physicalstu} in the Mandelstam plane.

Instead of $s,t,u$, we also use
\beq\label{eq:varEEW}
p q/M_K \eq \Eg ~,\quad p p'/M_K \eq \Ep~,\quad 
W^2 \eq (p_l+p_\nu)^2~,
\eeq
where $\Eg,~\Ep$ are the photon and the pion energy 
in the kaon rest frame. This set is useful when 
discussing partial decay widths. 

In the case of four body decays we have five 
independent variables, thus two more variables are needed to describe fully
the kinematics of $\kl3g$ decays. We choose
\beq
p p_e/M_K \eq \Ee ~,\quad x \eq p_e q/M_K^2,
\label{var2}
\eeq
where $\Ee$ is the positron energy in the kaon rest frame.
The dimensionless variable $x$ is related to the angle
$\te$ between the photon and the positron:
\beq
x M_K^2 \eq \Eg \,\left( \Ee -
\sqrt{\Ee^2-m_e^2}\,\cos\te \right).
\label{theta}
\eeq

The total decay rate is given by
\begin{align}
&\Gamma(K^0 \to \pi^- e^+ \nu \gamma) \eq \label{rate}\\
& \hskip 1cm
\frac{1}{2 M_K (2\pi)^8}\,\int
d_{\rm LIPS}(p;\,p',p_e,p_\nu,q) \, \sum_{\rm spins} \left|T\right|^2 ~, \no
\end{align}
where $T$ is the amplitude in \eqref{me_Kpg}, 
and we denote
the Lorentz invariant phase space element for
the $\kl3g$ process
by $d_{\rm LIPS}(p;\,p',p_e,p_\nu,q)$.\footnote{For the decay 
of a particle of momentum $p$ into $n$
  particles of momenta $p_1,\dots,p_n$, one has
\[
d_{\rm LIPS}(p;p_1,\dots,p_n) \eq
 \delta^4\Bigl(p-\sum_{i=1}^n p_i\Bigr)\prod_{k=1}^n\frac{d^3 p_k}{2p_k^0}~.
\]}
The square of the matrix element \eqref{me_Kpg}, summed over photon
and lepton polarizations, is a bilinear form in the invariant amplitudes 
$V_i,A_i, f_+$ and $f_2$. Performing the traces over the spins, we work with 
massless spinors, as a result of which the form factors
$A_3,V_3$ and $f_2$ drop out in the final expressions. 
[The electron mass cannot be set to zero everywhere, because the IB part 
  of the transition amplitude contains  mass singularities,
  generated by the diagram Fig.~\ref{fig:diag_K0}b.] 
In Appendix~\ref{app:traces}, we display the explicit result for 
$\sum_{\rm spins} \left|T\right|^2$, in particular the $T$-odd terms that are
generated by the imaginary parts of the structure functions, and comment on
the relation to the width of $K_L$.

\section{Analytical results from ChPT\label{sec:chpt}}

While Low's theorem furnishes a recipe to evaluate the terms of order $1/q$
and $q^0$ of an amplitude associated with a general radiative process, 
it does not give any insight into the terms of order $q$ and higher, 
that is the SD part. 
A convenient tool to derive expressions for the SD amplitude is ChPT. 
For the axial part, ChPT directly generates the corresponding amplitude in
a series expansion in the momenta, the leading contribution is generated
by the Wess-Zumino-Witten (WZW) term~\cite{wzw}. 
As for the vector amplitude, the chiral expansion contains both IB and SD
terms, hence  
if one simultaneously evaluates the $K_{l3}$ matrix element, 
the decomposition \eqref{eq:VIB} leads to the chiral expansion of the SD term.

\subsection{ChPT results at order \boldmath{$p^4$}}

In \cite{BEG}, the chiral expansion was carried out up to $\Order(p^4)$
for the neutral and for the charged decay modes.
[A tree-level calculation up to this order without the loop
contributions was performed in~\cite{holstein}.]
We do not describe the
calculation here and refer the interested reader to the original article.
The result for the SD terms is as follows. For the axial amplitude, 
one has
\beq \label{axial}
A_2 = -\frac{1}{8\pi^2F^2} ~,~~
A_1 = A_3 = A_4 = 0 ~~ [\Order(p^4)] ~. 
\eeq
$F$ is the pion decay constant in the chiral limit.\footnote{Usually,
  the meson decay constant in the SU(3) chiral limit is denoted by
  $F_0$. We refrain from following this convention in order to
  slightly ease the notation.}
We display the result for the vector amplitude $V_{\mu\nu}^{\rm SD}$ in 
terms of the Lorentz invariant form factors $V_i$,
\beq\begin{split}
V_1 &= \sqrt{2}\,\tilde I_2 ~,\\
V_2 &= \frac{\sqrt{2}}{qW}\left(\tilde I_1-p'q\,\tilde I_2+\sqrt{2}\triangle
f_+\right) ~,\\
V_3 &=\frac{\sqrt{2}}{qW}\left(\tilde I_3-\tilde f_2^+(W^2)\right) ~,\\
V_4 &= 0 \hspace{1cm} [\Order(p^4)]~.  \label{chiralVi}
\end{split} \eeq
The integrals $\tilde I_i,\,\tilde f_2^+$ are defined as follows. In
\cite{BEG}, the  
one-loop expression for $V_{\mu\nu}$ in the charged decay
mode $K^+_{l3\gamma}$ is defined in terms of integrals $I_i,\,f_i^+$,
explicitly displayed there.
 The quantities $\tilde
I_i,\,\tilde f_2^+$ are obtained from $I_i,\,f_2^+$ by
\begin{enumerate}
\item replacing the arguments $(p,p')$ by $ -(p',p)$ in the $I_i\,$; 
\item inserting the appropriate coefficients $c_i^I$ for
  $K^0_{l3\gamma}$ listed in Table 10 of that reference. 
\end{enumerate}
Note in addition that $\triangle f_+$ in \eqref{chiralVi} refers to the
chiral one-loop representation of this quantity.

It turns out that the form factors
$V_i$ are nearly constant over the physical phase space. 
This is due to
the fact that in the vicinity of the physical phase space, there are no
singularities at this order in the chiral expansion. This fact allows one to
derive expressions for the SD parts that are considerably simpler than the full
ones, yet still precise enough for our purpose. In a first step, we expand
these amplitudes in powers of the photon momentum $q$ and  keep only the
leading order term. This amounts to setting $s=M_\pi^2,u=M_K^2$, as a result
of which the $V_i$ become functions of $t$ alone. It turns out that all loop
integrals can be expressed in terms of the standard one-loop integral $\bar
J(t)$. The resulting expressions are displayed in Appendix~\ref{app:Vi}.
An even more drastic simplification results when one furthermore sets $t=0$ in
the simplified formulae. The result reads
\begin{align}
V_1 &= -\frac{8}{F^2}\bar{L}_9 - \frac{(1-x)^{-2}}{32\pi^2F^2} \biggl\{ 
  \frac{1}{3}\left(53-25x+2x^2\right) 
\no\\ & \hskip 0.45cm
+ \left(1+x-x^2+x^3\right)\frac{\log x}{2(1-x)} 
\no\\ & \hskip 0.45cm
- \left(127-93x+21x^2-x^3\right)\frac{\log y}{2(1-x)} \biggr\}
+\Order(q,t) ~, \no\\ 
V_2 &= \frac{4}{F^2} \left( \bar{L}_9 +\bar{L}_{10} \right)
\no\\ & \hskip 0.45cm 
+ \frac{(1+x)(1-x)^{-2}}{64\pi^2F^2} \biggl\{ 
  1+x + \frac{2x \log x}{1-x}  \biggr\} 
\no\\ & \hskip 0.45cm 
- \frac{(1-x)^{-3}}{32\pi^2F^2} \biggl\{ 
  \frac{166}{3}(9-4x)+(77-x)\frac{x^2}{3}
\no\\ & \hskip 0.45cm
+ x(3+2x) \frac{\log x}{1-x} 
- 9(12-x)(4-x)^2 \frac{\log y}{1-x}  \biggr\} 
\no\\ & \hskip 0.45cm
+\Order(q,t) ~, 
\no\\
V_3 &= - \frac{(1-x)^{-4}}{32\pi^2F^2\mk} \biggl\{ 
  \frac{2611}{3}-13x(34-5x)-\frac{4}{3}x^3 
\no\\ & \hskip 0.45cm
+ x(2+3x+x^2) \frac{\log x}{1-x} 
- 27(7-x)(4-x)^2 \frac{\log y}{1-x}  \biggr\} 
\no\\ & \hskip 0.45cm
+\Order(q,t) ~.\label{Vit=0}
\end{align}
Here, $x=M_\pi^2/M_K^2$, $y=M_\eta^2/M_K^2$. Furthermore, it is understood
that $M_\eta^2$ is related to $M_K^2,M_\pi^2$   through the Gell-Mann--Okubo
 relation.

\subsection{ChPT results at order \boldmath{$p^6$}\label{sec:p6}}

There are two main reasons to consider $\Order(p^6)$ corrections to
the structure-dependent terms as described in the previous subsection:
\begin{enumerate}
\item As the structure-dependent terms vanish at tree level in the chiral
  expansion, the one-loop or $\Order(p^4)$ predictions are only the leading
  order results for these amplitudes.  Subleading corrections are often
  sizeable in chiral SU(3), therefore it is mandatory to investigate
  $\Order(p^6)$ terms in order to be sure to control the size of the 
  structure-dependent terms.
  Furthermore, several of the structure functions vanish at
  leading (one-loop) order ($V_4$, $A_1$, $A_3$, $A_4$) or nearly
  so in the sense that they do not allow for natural-size counterterms
  ($V_3$), so the size of corrections to these is completely unknown from
  the one-loop results.
\item At $\Order(p^4)$, all structure functions are real in the physical
  region, and the cuts in these functions lie far outside the kinematically
  allowed range.  This is the reason
  why they are so smooth and can be approximated to such high accuracy by
  simple polynomials.  However, this changes at $\Order(p^6)$, as will be
  seen below.
\end{enumerate}
For the following discussion, we again use the Mandelstam variables
$s$, $t$, $u$.  The lowest-lying cuts for the structure functions in terms of
these three variables are as follows:
\begin{enumerate}
\item For the weak vector current, they start at $s_{\rm thr}=9M_\pi^2$, 
   $t_{\rm thr}=(M_K+M_\pi)^2$, $u_{\rm thr}=(M_K+2M_\pi)^2$,
   respectively.  Only the
   $t$-channel cut exists at $\Order(p^4)$ as the other two require 
   three-particle intermediate states and therefore occur only at two-loop
   order.
\item For the weak axial current, cuts start at  $s_{\rm thr}=4M_\pi^2$, 
   $t_{\rm thr}=(M_K+M_\pi)^2$, $u_{\rm thr}=(M_K+M_\pi)^2$,
   respectively.  All these occur
   at one-loop order, but are suppressed to $\Order(p^6)$ as they require
   an anomalous vertex.
\end{enumerate}
\begin{figure}
\vskip 2mm
\centering
\includegraphics[width=8.6cm]{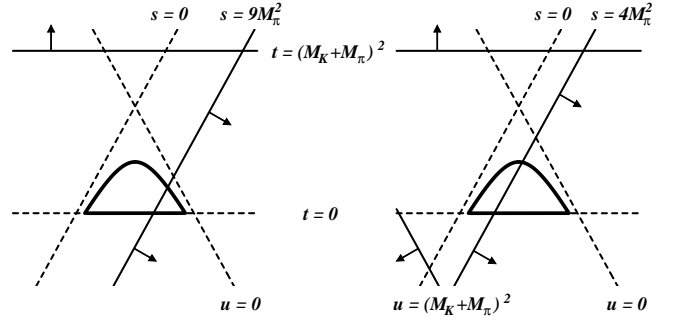}
\caption{Cuts in the complex plane for the vector (left panel) and axial
    (right panel) amplitudes for fixed $W^2=m_e^2$. The arrows indicate the
    part of the plane where the decay amplitude is complex, starting from the
    lowest cuts indicated by the full lines.  The limits of physical phase
    space are also shown (thick solid line). \label{fig:VAcuts}} 
\end{figure}
These cuts are displayed graphically in Fig.~\ref{fig:VAcuts}, where we have
drawn them in the Mandelstam plane together with the allowed phase space
for fixed (and minimal) $W^2=m_e^2$, which corresponds to the maximal range in
$s$, $t$, $u$.  While the $t$- and $u$-channel cuts lie far outside the
physical region, we note that the $s$-channel cuts overlap with it 
(precisely: for $W^2<(M_K-3M_\pi)^2$ in the vector and $W^2<(M_K-2M_\pi)^2$ in
the axial case), such that at least some of the structure functions become
complex at $\Order(p^6)$. 

\begin{figure}
\vskip 2mm
\centering
\includegraphics[width=8.5cm]{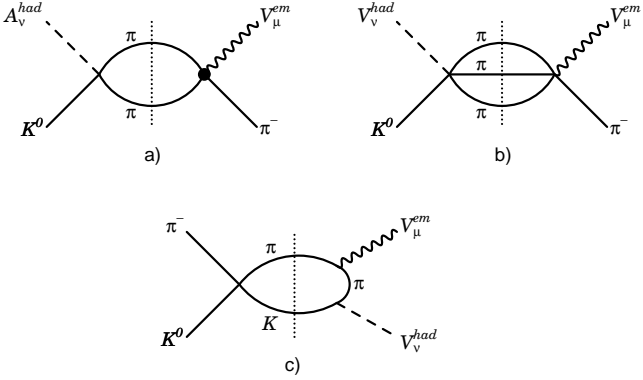} 
\caption{Feynman diagrams with cuts for a) $s>4\mpi$, b) $s>9\mpi$,
  c) $t>(M_K+M_\pi)^2$, respectively.  
  The filled vertex denotes a contribution from the anomalous
  Lagrangian at order $p^4$.
  The first two diagrams provide the
  structure functions with imaginary parts in the physical region,
  while the last one generates the only cut present in the amplitudes at
  $\Order(p^4)$.\label{fig:cutdiag}} 
\end{figure}
The diagrams with cuts responsible for imaginary \linebreak parts in the physical region
are displayed in Fig.~\ref{fig:cutdiag}, together with one typical diagram for the
$t$-channel cut appearing at $\Order(p^4)$.
Due to the smallness of phase space for the three-pion intermediate state, we
expect the effect of the cut in the vector structure functions to be tiny.  

For the above considerations, we have regarded $W^2$ as a fixed ``mass
squared'' of
the lepton-neutrino pair, which is of course not true.  There are additional
cuts in $W^2$, the lowest one starting at $W^2=(M_K+M_\pi)^2$ present at one
loop, but sufficiently far outside the physical region, plus a pole at
$W^2=M_K^2$ that however only appears in the structure function $A_3$.
The pole at $s=M_\pi^2$ defines the bremsstrahlung
part and is not present in the structure-dependent terms as defined by our
convention.

\subsubsection{Complete order $p^6$ corrections to the axial
  amplitudes\label{sec:p6axial}}

We have calculated the complete $\Order(p^6)$ corrections to
the axial structure functions $A_1$, $A_2$, and $A_4$. 
$A_3$ is always suppressed by a factor of $m_l^2$ and is therefore
disregarded in the context of the electron channel. 
We find the following structures:
\begin{align}
A_1 &= -\frac{1}{4\pi^2F_\pi F_K} \bigl\{
 S_1(s) + T_1(t) + U_1(u) +X_1  \bigr\}~, \label{A1struct} \\
A_2 &= -\frac{1}{8\pi^2F_\pi F_K} \bigl\{ 1+ 
  S_2(s)+T_2(t) + U_2(u) +X_2 \bigr\} ~, \label{A2struct}\\
A_4 &= -\frac{C_{4A}}{F_\pi F_K}~.\label{A4struct}
\end{align}
The explicit forms for the various loop functions as well as the
combinations of low-energy constants entering the expressions
\eqref{A1struct}-\eqref{A4struct} can be found in
Appendix~\ref{app:axial}. 
We remark that it is strictly necessary to differentiate
between $F_\pi$ and $F_K$ at this order only for $A_2$.
The normalization was chosen this way such that any dependence on the
low-energy constants $L_4$, $L_5$ vanishes in the final result.

We furthermore note that only $A_1$ has a contribution from the $s$-channel
cut and therefore becomes complex in parts of the physical region at this
order, while $S_2(s)$ is a simple polynomial.
It turns out, though, that also in $S_1(s)$ the standard
two-point loop function $\bar{J}_{\pi\pi}(s)$ from the two-pion intermediate
state comes with a prefactor of $(s-4M_\pi^2)$ such that the cusp in the real
part is smoothed out.
This is due to the fact that the $\pi\pi \to \pi \gamma$ rescattering 
has to be a $P$-wave and is therefore suppressed at threshold.
For the same reason, the imaginary part also rises very slowly above threshold:
as the leading and next-to-leading order
amplitudes in the chiral expansion are real, it is certainly
negligible in the squared matrix element at our accuracy. 

We conclude that even for the axial structure functions, the impact of the
various cuts on the behavior in the physical region is rather weak.

\subsubsection{Order $p^6$ corrections to the vector
               amplitudes\label{sec:p6vector}} 
A complete evaluation of the vector amplitudes at order $p^6$ requires a full
two-loop calculation, which is beyond the scope of this article. A less
ambitious work consists in the determination of the 
leading chiral logarithms  at two-loop order~\cite{doublelogs}.
As a first step in this direction,  we 
have explicitly calculated the contributions of the
form $L_i \!\times\! L_j$ at order $p^6$. We find that these can all be 
absorbed into a renormalization of the couplings at order $p^4$,
according to $F^2 \to F_\pi F_K$.  
 This is strictly analogous to the observation 
that the dependence on $L_4$, $L_5$ for the axial
terms can be absorbed into such a renormalization.
We will make use of this
fact in Sect.~\ref{sec:sdnum}, where we provide numerical values for the SD
terms. In addition, we will give a rough estimate of the
contributions at order $p^6$, and defer an evaluation of the leading  
logarithms to a later publication~\cite{inprogress}.

\section[The ratio           $\rR$ ]
        {The ratio \boldmath{$\rR$}}\label{sec:r}

A particularly useful quantity to consider for the examination
of $K_{e3\gamma}$ decays is the ratio $\rR$ defined in \eqref{def:r},
rather than the absolute width for the $K_{e3\gamma}$ channel or the 
branching ratio thereof.  
This is desirable both from the experimental and the theoretical point of view
for rather similar reasons:  both experimentally and theoretically,
certain normalization factors cancel in the ratio (to a large extent),
and hence the uncertainties ensuing thereof are avoided.
To present the situation in a more transparent way, we shall initially neglect
all possible complications ensuing from radiative corrections or isospin
breaking, and shall discuss these afterwards in Sect.~\ref{sec:R}.
In order to remind the reader of this simplification, we shall denote $\rR$ in
the absence of real and virtual photon corrections by $\rr$ in the following,
\beq \alpha^{-1} \rr \eq \left[\alpha^{-1} \rR\right]_{\alpha=0} ~. \eeq
We may decompose $\rr$ according to $\rr = \rr^{\rm IB}+\rr^{\rm SD}$
in the following sense: 
$\rr^{\rm IB}$ is 
understood to be $\rr$ in the limit where all structure-dependent 
terms are omitted, while we may then define
$\rr^{\rm SD} = \rr-\rr^{\rm IB}$, such that $\rr^{\rm SD}$ contains
in particular also interference terms of bremsstrahlung and 
structure-dependent terms.

We begin by deriving a simple expression for $\rr$.
Starting from \eqref{rate}, we may define a quantity $\ol{\rm SM}$ by
\beq \begin{split}
\Gamma(K_{e3\gamma}) 
&= \frac{1}{2M_K(2\pi)^8} \int d_{\rm LIPS} \sum_{\rm spins} |T|^2 
\\ & \doteq 
  \frac{8\alpha M_K^5 G_F^2 |V_{us}|^2}{(2\pi)^7} f_+(0)^2
   \int d_{\rm LIPS} \; \ol{\rm SM} ~. \label{def:SM}
\end{split}
\eeq
The phase space integral $\int d_{\rm LIPS} \;\ol{\rm SM}$ such defined is
dimensionless and free of (electroweak) coupling constants.  
For the bremsstrahlung part, also $f_+(0)^2$
factors out naturally, such that all form factors appearing 
in $\ol{\rm SM}$ are the normalized form factors $\bar{f_+}(t)=f_+(t)/f_+(0)$.
The non-radiative width can be written as
\beq \label{eq:factorf+}\begin{split}
\Gamma(K_{e3}) &= \int dy \, dz \,\rho(y,z) \\
\rho(y,z) &= 
\frac{M_K^5 G_F^2 |V_{us}|^2}{128\pi^3} f_+(0)^2 A(y,z) \bar{f}_+(t)^2 
\end{split}\eeq
where $y=2pp_e/M_K^2$, $z=2pp'/M_K^2$, and 
\beq \begin{split}
A(y,z)&\eq 4(z+y-1)(1-y)+r_e(4y+3z-3) \\ & \hskip 0.45cm
-4r_\pi+r_e(r_\pi-r_e) ~,
\end{split} \eeq
with $r_e=m_e^2/M_K^2$, $r_\pi=M_\pi^2/M_K^2$.
We therefore find for $\rr$ the following simple expression,
\beq
\rr = \frac{8\alpha}{\pi^4} \frac{\int d_{\rm LIPS} \, \ol{\rm SM}}
                               {\int dy\,dz\,A(y,z)\bar{f}_+(t)^2} 
~, \label{r}
\eeq
in which all factors $G_F$, $V_{us}$, $f_+(0)$, 
and $M_K$ have canceled. [For the relation between $K_{e3\gamma}$ and
$K_{e3\gamma}^0$ decays, see Appendix~\ref{app:traces}.] 

\subsection{Phase space integrals}

Assuming
\beq
\bar{f}_+(t) = 1+\lambda_+ \frac{t}{M_\pi^2}+\lambda''_+ \frac{t^2}{M_\pi^4} ~,
\eeq
one may expand the integral in the denominator according to 
\begin{align} 
I &\eq \int dy\,dz\,A(y,z)\bar{f}_+(t)^2 \label{kl3density} \\
&\eq a_0 + a_1 \lambda_+ + a_2 \left(\lambda_+^2+2\lambda''_+\right)
+ a_3 \lambda_+ \lambda''_+ + a_4 {\lambda''_+}^2 ~. \no
\end{align}
The numerical values for the coefficients $a_{0-4}$ as found by performing
the relevant phase space integrals are given in
Table~\ref{tab:ai}. 
\begin{table}
\centering
\caption{Coefficients for the $K_{e3}$ phase space  integral.\label{tab:ai}}
\medskip
\renewcommand{\arraystretch}{1.4}
\begin{tabular}{ccccc}
\hline
$a_0$ & $a_1$ & $a_2$ & $a_3$ & $a_4$ \\
\hline
0.09390 & 0.3245 & 0.4485 & 3.092 & 6.073 \\
\hline
\end{tabular}
\renewcommand{\arraystretch}{1.0}
\end{table}
We remark that although we neglect isospin breaking effects in this subsection,
we have employed the physical kaon and pion masses 
[Appendix~\ref{app:notation}].

Similarly, one can also calculate the dependence of the numerator on the form
factor parameters $\lambda_+$, $\lambda''_+$.  In an analogous manner to
\eqref{kl3density} we write
\begin{align} 
I^\gamma &\eq \int d_{\rm LIPS} \, \ol{\rm SM} \label{kl3gdensity} \\
&\eq b_0 + b_1 \lambda_+ + b_2 \lambda_+^2+ b_3 \lambda''_+
+ b_4 \lambda_+ \lambda''_+ + b_5 {\lambda''_+}^2 ~. \no
\end{align}
The coefficients $b_0$, $b_1$, and $b_3$ have contributions
also from the structure-dependent terms, such that we decompose them again
according to $b_i=b_i^{\rm IB}+b_i^{\rm SD}$.  $b_2$, $b_4$, and $b_5$ have no
structure-dependent part. 
In our framework, the bremsstrahlung amplitude is expressed in terms of
a completely general (phenomenological) form factor $f_+(t)$, \linebreak
while the coefficients $b_i^{\rm SD}$ can be chirally expanded and receive
their leading contribution at $\Order(p^4)$.
We point out that all the coefficients $b_i$ depend on the
experimental cuts $\Ecut$, $\tecut$,
we however suppress this dependence in our notation.

We mention that, in principle, the inclusion of struc\-ture-dependent terms  
re-introduces a dependence on \linebreak $f_+(0)$ by which these terms have to be divided
in order to arrive at \eqref{r}.  However, the uncertainty
in the structure-dependent terms themselves coming from higher order
($\Order(p^6)$) contributions is at least one order of magnitude larger than
the uncertainty in $f_+(0)$, so we do not have to worry about a very precise
value for the latter.  
For our purposes, we have used the value predicted (parameter-free)
in one-loop ChPT, $f_+(0)=0.977$~\cite{leutroos}.

\begin{table}
\centering
\caption{Coefficients for the $K_{e3\gamma}$ phase space integral
for  $\Ecut=30$~MeV, $\tecut=20^\circ$.
The errors for $b_i^{\rm SD}$ are $p^6$ estimates.
\label{tab:bi}
}
\renewcommand{\arraystretch}{1.4}
\begin{tabular}{cccccc}
\hline
$b_0^{\rm IB}$ &  $b_1^{\rm IB}$ & $b_2$ & $b^{\rm IB}_3$ &  $b_4$ & $b_5$ \\
\hline
1.509 &  5.23 &  6.92 & 14.71 &  47.6 & 92.3 \\
\hline
\mc{$b_0^{\rm SD}$} & \mc{$b_1^{\rm SD}$} & \mc{$b^{\rm SD}_3$}  \\
\hline
\mc{$-0.011\pm0.003$} & \mc{$-0.02\pm0.01$} & \mc{$ -0.06\pm0.02$}  \\
\hline
\end{tabular}
\renewcommand{\arraystretch}{1.0}
\end{table}
The values for the coefficients $b_i$ can only be found numerically in this
case.  Our findings for the ``standard cuts'' $\Ecut=30$~MeV,  
$\tecut = 20^\circ$ are collected in Table~\ref{tab:bi}.
For the values of the low-energy constants see Appendix~\ref{app:notation}.
We neglect any variation in these constants as we include estimates of
the uncertainties stemming from the $\Order(p^6)$ contributions 
(see Sect.~\ref{sec:p6})
that generously cover the range of values for $L_9^r$, $L_{10}^r$.  
These uncertainties are quoted as errors for the $b_i^{\rm SD}$ in
Table~\ref{tab:bi}.  
We describe the precise procedure how we estimate these ranges numerically
in Sect.~\ref{sec:Np6}
and only note for now that the possible corrections 
are roughly 30\%, which one would naively expect for chiral
SU(3).

The central observation here is that structure-depen\-dent terms as predicted
by ChPT at one loop contribute as little as 1\% to each of the parameters in
Table~\ref{tab:bi} and therefore to the total radiative decay rate.

\subsection{Form factor dependence of \boldmath{$\rr$}}

We are now in the position to give a numerical prediction for $\rr$ 
that depends solely on $\bar{\lambda}_+ = \lambda_+/\lambda_+^c$
and $\bar{\lambda}''_+=\lambda''_+/(\lambda_+^c)^2$. 
We choose to normalize all parameters by the central value
$\lambda_+^c=0.0294$, see Appendix~\ref{app:notation},
in order to expand in terms of quantities of natural order of magnitude. 
Note that ${\lambda_+^c}^2$ is a natural scale for $\lambda''_+$ that
one would obtain e.g.\ from $K^*$ dominance.

The numerical prediction is obtained from 
\eqref{r}, \eqref{kl3density}, and \eqref{kl3gdensity}.
In order to make the form factor dependence more transparent, though, we
expand $\rr$ according to 
\begin{align}
\rr\left(\bar{\lambda}_+,\bar{\lambda}''_+\right)
\eq \rr(1,0) \Bigl\{ 1
&+ c_1 \, \left(\bar{\lambda}_+ -1 \right) 
+ c_2 \, \left(\bar{\lambda}_+ -1 \right)^2
\no\\ & 
+ c_3 \, \bar{\lambda}''_+
 + \ldots \Bigr\} ~, \label{Nr}
\end{align}
where we only retain the leading (and most important) terms.

We begin again by considering bremsstrahlung only.
The numerical results are given in Table~\ref{tab:ci}, 
again for the standard cuts.
They demonstrate that $\rr^{\rm IB}$ is \emph{extremely}
insensitive to the details of the $K_{e3}$ form factor due to a large
cancellation of the $\lambda_+$ ($\lambda''_+$) dependence in numerator and denominator
of $\rr^{\rm IB}$.\footnote{The weaker $\lambda_+$ dependence of $\rr$
was already hinted at in a footnote in~\cite{doncel}.} 
Furthermore, we note that a tree-level calculation of $\rr$ in ChPT would amount
to $\rr^{\rm IB}$ (as there are no structure-dependent terms) with 
$\bar\lambda_+=\bar\lambda''_+=0$ (point-like form factors).  
Numerically, one finds $\rr^{\rm tree}=0.963\times 10^{-2}$,
which is even identical to the above result in all digits
displayed.

Inclusion of the structure-dependent terms is straightforward from the results
given in Table~\ref{tab:bi}, we show the numerical results for the complete 
(IB+SD) coefficients also in Table~\ref{tab:ci}. 
Perpetuating what was done in Table~\ref{tab:bi}, we again quote errors on all
parameters as induced by the estimated $\Order(p^6)$ uncertainties.

We repeat here the observation made in the previous subsection that 
structure-dependent terms contribute as little as 1\% to the ratio $\rr$.
In view of the above remark about $\rr$ at tree
level, the complete one-loop correction to $\rr$ is in fact about 1\%.  
Or, to put it even differently, a prediction of the radiative decay rate based 
solely on inner bremsstrahlung is expected to have a precision of about 1\%.
The parameters $c_i$ are shifted more visibly due to the fine cancellation
between numerator and denominator of $\rr$, but they remain tiny and do not
change the conclusion that the form factor dependence of $\rr$ is entirely
negligible.  

\begin{table}
\centering
\caption{Coefficients for the $\bar{\lambda}_+$, $\bar{\lambda}''_+$ 
  dependence of $\rr^{\rm IB}$, $\rr$.  
  The errors for the $c_i$ are $p^6$ estimates.
  All numbers are given for the standard cuts.\label{tab:ci}
}
\medskip
\renewcommand{\arraystretch}{1.4}
\begin{tabular}{cccc}
\hline
$\rr^{\rm IB}(1,0)\times 10^2$ & $c_1^{\rm IB}\times 10^3$ 
& $c_2^{\rm IB}\times 10^4$ & $c_3^{\rm IB}\times 10^4$  \\
\hline
$0.963$ & $-0.0$ & $ -1.5 $ & $ 1.2 $ \\
\hline
$\rr(1,0)\times 10^2$ & $c_1\times 10^3$ & $c_2\times 10^4$ & $c_3\times 10^4$  \\
\hline
$0.957\pm0.002$ & $0.3\pm0.2$ & $-1.6\pm0.2$ & $1.5\pm0.2$ \\
\hline
\end{tabular}
\renewcommand{\arraystretch}{1.0}
\end{table}

This is an appropriate place to compare our 
findings with the calculation in~\cite{BEG}. 
We note that there, the branching ratio BR$(K_{e3\gamma})$ 
was determined from
the chiral amplitude at order $p^4$, with $\Ecut= 30$ MeV, $\tecut =
20^\circ$. As is clear from the above, a
cancellation of the momentum dependence of the form factors does not occur
in this case. In order to compare with the present calculation, we 
use the formula \eqref{r} and note that the value 
for $L_9^r$ used in~\cite{BEG}
corresponds to $\lambda_+=0.032$. We have repeated the calculation  
with the matrix element at order $p^4$ 
provided in \cite{BEG}. By use of  \eqref{r}  
 and \eqref{kl3density}, we find
$\rr=0.96\times10^{-2}$, in perfect agreement with the value displayed in
the fourth row in Table~\ref{tab:ci}. 

The important conclusion is that imprecise knowledge of the
$K_{e3}$ form factor does not preclude a precise prediction of $\rr$.

\subsection{Dependence on the experimental cuts}

The near-complete cancellation of all form factor dependence in $\rr$
suggests the question whether this might be accidental due to the
specific cuts chosen for the radiative decay width.
Here, we want to briefly analyze how the above findings change when
we vary the cuts on $\Eg$ and $\te$.  We restrict
ourselves to the bremsstrahlung part of the radiative width and $\rr$.
The most important information on the expansion of $\rr$ according to
\eqref{Nr} is collected in Table~\ref{tab:ciIBcuts}.
We find that the coefficients $c_i^{\rm IB}$ do indeed vary considerably
for different cuts, but always stay ``small'', 
$c_1^{\rm IB}=\Order(10^{-3})$, $c_{2/3}^{\rm IB}=\Order(10^{-4})$.
The suppression of $c_1^{\rm IB}$ far beyond $10^{-3}$ for the standard cuts
turns out to be accidental, however.  
Still, with $|\bar{\lambda}_+ -1| \lesssim 0.1$,
$|\bar{\lambda}''_+| \sim 1$, we find that the $K_{e3}$ form factor
affects $\rr$ at the level of $10^{-4}$.

We should remark here on the latest results for these form factor parameters
published by the KTeV Collaboration~\cite{ktevlambda}.  
For the first time, they find 
significant statistical evidence for a non-zero quadratic term in $f_+(t)$,
together with a sizeable reduction of $\lambda_+$.
Converted to our conventions, the combination of
their quadratic fits to $K_{e3}$ and $K_{\mu 3}$ corresponds to
$\bar{\lambda}_+ = 0.70 \pm 0.06$,  
$\bar{\lambda}''_+ = 1.85 \pm 0.40$.\footnote{Similar
  trends were already noted in the theoretical fits in~\cite{btkl3}.
  Note however the latest experimental results from NA48~\cite{Lai},
  where a free quadratic fit leads to $\bar{\lambda}_+=0.95\pm 0.08$,
  $\bar{\lambda}''_+=0.23\pm 0.52$, completely consistent with our
  central values.} 
While the deviation for the two individual parameters is quite sizeable, we
find from Table~\ref{tab:ciIBcuts} that a simultaneous reduction of
$\bar{\lambda}_+$ and an enhancement in $\bar{\lambda}''_+$ still only modify
$\rr$ at the order of a few parts times $10^{-4}$ at best.
Our finding that $\rr$ is independent of the details of $f_+(t)$ to a very
large extent therefore remains valid.
\begin{table}
\centering
\caption{Coefficients for the $\bar{\lambda}_+$, $\bar{\lambda}''_+$
  dependence of $\rr^{\rm IB}$ with variation of the experimental cuts
  on $\Ecut$,  $\tecut$. \label{tab:ciIBcuts}
}
\medskip
\renewcommand{\arraystretch}{1.4}
\begin{tabular}{cccccc}
\hline
$\!\Ecut\!$ & $\!\tecut\!\!$ & 
$\!\rr^{\rm IB}(1,0)\!\times\! 10^2\!\!$ & 
$\!c_1^{\rm IB}     \!\times\! 10^3\!\!$ & 
$\!c_2^{\rm IB}     \!\times\! 10^4\!\!$ & 
$\!c_3^{\rm IB}     \!\times\! 10^4\!\!$  \\
\hline 
\!20\,MeV\!\! & $\!20^\circ\!\!$ & 1.297 & $-2.1$ & $-0.4$ & $-2.0$ \\
\!30\,MeV\!\! & $\!20^\circ\!\!$ & 0.963 & $-0.0$ & $-1.5$ &   1.2  \\
\!40\,MeV\!\! & $\!20^\circ\!\!$ & 0.743 &   2.1  & $-2.6$ &   4.5  \\
\hline
\!30\,MeV\!\! & $\!10^\circ\!\!$ & 1.254 &   1.7  & $-1.9$ &   3.3  \\
\!30\,MeV\!\! & $\!20^\circ\!\!$ & 0.963 & $-0.0$ & $-1.5$ &   1.2  \\
\!30\,MeV\!\! & $\!30^\circ\!\!$ & 0.790 & $-1.6$ & $-1.1$ & $-0.7$ \\
\hline 
\end{tabular}
\renewcommand{\arraystretch}{1.0}
\end{table}

\subsection{Isospin breaking}\label{sec:R}

We have seen above that the ratio $\rr$ can be 
predicted to an amazingly good precision of less than 1\%,
using ChPT to one loop for the structure-dependent terms
plus a rough estimate of the size of higher-order corrections. 
At this level of precision, isospin breaking corrections -- generated by real
and virtual
photons, and by $m_u-m_d\neq 0$ -- become  relevant, and we
discuss  these  here.

 As soon as one
includes virtual photon corrections, one also has to take care of additional
soft photon radiation in order to obtain an infrared finite quantity, and
 we therefore clarify  the precise prescription as to what is meant by
the numerator and the denominator of the ratio $\rR$ in  \eqref{def:r}.
In accord with the  experimental situation~\cite{ktev,na48}, 
 the denominator denotes
the \emph{inclusive} width for $K_L \to \pi^\pm e^\mp \nu_e (n
\gamma)$,
where $(n \gamma)$ denotes any number of photons of arbitrary energy.  
The numerator is specified in an analogous manner: experimental
measurements of $K_{e3\gamma}$ require detection of at least one
hard ($\Eg > \Ecut,\, \te > \tecut$) photon in the final state, 
plus an arbitrary number of additional soft or hard photons.
A full calculation of the $\Order(\alpha^2)$ contributions in $\rR$
is beyond the scope of this work. Instead, we identify some partial 
contributions to it and give an estimate of
the remainder. 

Radiative corrections have been evaluated 
for $K_{e3}$ in \cite{kl3rad,bitev,Andre}. Effects from the quark mass
difference $m_u-m_d$ have been in addition 
taken into account in~\cite{kl3rad}.
We note the following from that investigation:
\begin{enumerate}
\item One particularly pronounced effect is the electroweak correction
factor to the Fermi constant, $G_F^2 \!\to\! S_{EW}G_F^2$, which contains
a large short distance enhancement factor~\cite{marcianosirlin,kl3rad} 
$\propto \log M_Z/M_\rho$ such that $S_{EW}-1 \approx 2.2\times 10^{-2}$. 
This factor, however, is universal in the sense that it applies
identically also to the radiative rate and therefore cancels in $\rR$. 
\item There are electromagnetic vertex corrections and $m_u-m_d$ effects 
that are collected in a shift in $f_+(0)$, which 
can still be factored out as in \eqref{eq:factorf+}.
 The remaining corrections have been incorporated in an expansion
of the integral $I$ in terms of
$K_{e3}$ form factor parameters according to \eqref{kl3density}.
The authors of~\cite{kl3rad} have calculated the values 
for all the parameters $a_{0-4}$ including corrections of 
$\Order(e^2 p^2, (m_u-m_d)p^2$).  
Their results are collected in Table~\ref{tab:aiem}.\footnote{We are grateful 
  to the authors of~\cite{kl3rad} for providing us with the values for
  $a_3$ and $a_4$ which are not included in the publication.}
\begin{table}
\centering
\caption{Coefficients for the $K_{e3}$ phase space integral,
including corrections of $\Order(\alpha, m_u-m_d)$.
The numbers for $a_1$ to $a_4$ are taken from~\cite{kl3rad}.  
For $a_0$, see main text. 
\label{tab:aiem}
}
\medskip
\renewcommand{\arraystretch}{1.4}
\begin{tabular}{ccccc}
\hline
$a_0$ & $a_1$ & $a_2$ & $a_3$ & $a_4$ \\
\hline
0.09412 & 0.3241 & 0.4475 & 3.080 & 6.042 \\
\hline
\end{tabular}
\renewcommand{\arraystretch}{1.0}
\end{table}
In~\cite{kl3rad}, the corrections from real photon emission
were treated slightly differently from what is done here:  while
there was no upper cut on the photon energy, the remaining phase space
integration of pion and electron momenta was restricted to the
kinematics compatible with $K_{e3}$ phase space.  
In order to agree with the experimental situation for the case at hand,
 we have removed this cut,
 and have modified $a_0$ accordingly, 
augmenting it by 0.57\%. 
\item It remains to estimate isospin breaking effects
in the numerator of $\rR$.
A source of potentially large radiative corrections  
are electron mass singularities. Because the observed photon is hard and 
emitted with an angle $\theta^*_{e\gamma} > 20^\circ$ with respect to the
electron, we expect that, according to the 
KLN~\cite{KLN} theorem, 
these may  be absorbed into a running  electromagnetic coupling 
constant.
 In the present case, the initial state is neutral -- we
therefore evaluate the running coupling  at the pion mass, rather than
the kaon mass,
 $\alpha\rightarrow\alpha(1+\frac{\alpha}{3\pi}\log(M_\pi^2/m_e^2))$. 
[We stick to corrections of relative order $\alpha$ here. 
 Evaluating the coupling instead at the
 kaon mass affects the final result for $R$ by about two permille.
 Up to the number of digits displayed below, the final number remains
 unchanged.] 
We denote the remaining relative
 corrections by $\Delta_{\rm em}$. 
We expect them to be  small, 
of the order $\alpha/\pi\simeq 2.3\times 10^{-3}$. To be on the safe
side, we increase this estimate by a factor five and take
$\Delta_{\rm em}=0.01$.
\item
Finally, we note that part of the running coupling is absorbed into $f_+(0)$
that contains, in the convention of~\cite{kl3rad}, an electron mass
singularity as well. We factorize this piece as before, such that the
effect of the mass singularity in the numerator is reduced, 
 $\alpha\rightarrow\alpha(1+\frac{\alpha}{12\pi}\log(M_\pi^2/m_e^2))$.
As for isospin breaking through \linebreak $m_u-m_d$, we expect 
the effects that cannot be absorbed into
$f_+(0)$ to be tiny, and we neglect them here.
\end{enumerate}

\subsection{Final result for \boldmath{$R$}\label{sec:finalR}}

Compared to results quoted in Table~\ref{tab:ci}, our prediction is therefore
modified by isospin breaking  corrections in the following manner: 
 the central value is
reduced by 0.2\% due to corrections in the denominator. We use in the
numerator the running coupling as discussed above, and add an
uncertainty of $\pm \Delta_{\rm em}$. 
We finally find
\beq
\rR = \left( 0.96 \pm 0.01 \right) \times 10^{-2} 
\eeq
as our prediction.
This may be compared to 
$\rr^{\rm IB} = 0.963\times 10^{-2}$ for bremsstrahlung only, without
radiative corrections.
\begin{figure}
\vskip 2mm
\centering
\includegraphics[width=7.5cm]{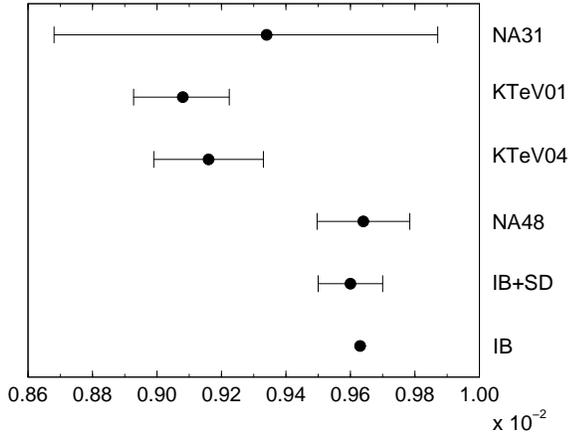} 
\caption{Our prediction for $\rR$ (dubbed ``IB+SD'') 
  and $\rr^{\rm IB}$ (``IB'') in comparison with
  experimental results for $ R$ 
   from NA31~\cite{leber}, KTeV~01~\cite{ktev},
  KTeV~04~\cite{ktev_04}, and NA48~\cite{na48}.  Note that $\rr^{\rm
  IB}$  does not contain radiative corrections.
  All values refer to the ``standard cuts'' $\Ecut=30\,$MeV, 
  $\tecut=20^\circ$.
  \label{fig:R}} 
\end{figure}
Both $\rR$ and, as a point of reference, $\rr^{\rm IB}$
are displayed in Fig.~\ref{fig:R}, together
with experimental results from \cite{leber,ktev,na48,ktev_04}.
Note  that the corresponding  Fig.~4
in~\cite{ktev} does not properly represent the theoretical result obtained in
ChPT in~\cite{BEG}: in that reference, the ratio 
$\Gamma(K_{e3\gamma})/\Gamma(K_L)$ was calculated, and an error analysis was 
not performed. In addition, the other two theoretical 
works~\cite{FFS,doncel} 
-- displayed with an error bar in that figure -- do also 
not contain an error estimate.

We conclude with the observation that the smallness of 
structure-dependent contributions in $\rR$ precludes a direct
determination of (hadronic) structure effects from $\rR$ alone.  
In order to extract such effects from experiment, one has to resort to
differential distributions, which we will 
discuss in Sect.~\ref{sec:distributions}.

\section[Structure-dependent terms: Num.\ results from ChPT]
        {Structure-dependent terms: \\ Numerical results from ChPT}
\label{sec:sdnum}

In the ratio $\rR$, the effect of the structure-dependent terms is tiny. 
On the other hand, in \cite{ktev}, the KTeV Collaboration 
attempted to extract two of the
structure-dependent terms from the $\Eg$ spectrum. Their result
encourages us to take up this issue here, in particular so in view of its
connection with the effective theory of the Standard Model.
 The remaining part of this article is devoted to this issue.

The structure-dependent terms are characterized by six amplitudes $V_i$,
$A_i$. The effect of  $V_3,A_3$ is suppressed by powers of the electron
mass in the decay rate -- these amplitudes 
 are only needed for a comparison with the basis used in \cite{ktev}.
 The $V_i, A_i$ are in general complicated functions of 
the variables $s$, $t$, and $u$. 
At leading order in ChPT, they are however real, vary  little over 
physical phase space, and may well be approximated
by real constants. It thus appears that in this approximation, the
amplitudes can directly be confronted with the KTeV analysis~\cite{ktev}.
 The reason why this is not the case is the following. As we have mentioned 
in Sect.~\ref{sec:kin}, the IB terms used in the present work
differ from the ones  in~\cite{FFS} [and used by the KTeV Collaboration], and
therefore the SD terms differ.
 In addition, we use a different set of tensors 
to decompose the SD terms into Lorentz invariant amplitudes.
 We have discussed before why we believe that the decomposition
 into IB and SD terms used here is more appropriate in $K_{e3\gamma}$ 
than the one originally proposed in~\cite{FFS}. As for the choice 
of a tensor basis, the one used here has the advantage that it
automatically singles out the two amplitudes whose effect in the 
rate is suppressed by powers
of the electron mass. The basis used in~\cite{FFS,ktev}
does not have this property, as a result of which the interpretation of the
various SD terms is somehow involved, see below.

We present the result of our analysis in the following manner. First, we
discuss numerical results for $V_i,A_i$ at leading and next-to-leading order in
ChPT. We then detail the decomposition of the hadronic tensors $V_{\mu\nu},
A_{\mu\nu}$ used in \cite{ktev}, and
translate our  result into the Lorentz invariant SD amplitudes 
used there.

\subsection{Numerical evaluation of the \boldmath{$\Order(p^4)$} terms\label{sec:Np4}}

An important assumption of the analysis of structure-dependent terms
in~\cite{ktev} is that these are real and constant -- which is not really
true. Let us therefore  investigate  
in what sense real and constant structure functions can  
be taken as reasonable approximations.

Whereas the leading contributions to the $V_i,\,A_i$ are real,
 \emph{imaginary parts} develop at higher orders in the chiral expansion.
 It is shown in Appendix~\ref{app:traces} that
 their effect is suppressed in the physical 
quantities considered in this work. More
 precisely, imaginary parts  occur only through contributions 
quadratic in the SD terms and are therefore completely  negligible here.
 Concerning the use of \emph{constant  form factors},  we have
 already mentioned that the leading contribution to the SD terms indeed
is very slowly varying over physical phase space. To quantify this statement,
we average the real part of the ChPT structure functions, 
i.e., integrate over phase space 
and divide by the phase space volume. We
quote the standard deviation for this average in order to quantify how
sensible the assumption of the functions being constant is.  
We use the notation $\av{V_i}$  for the result of this averaging procedure,
 and quote the numbers in units of the kaon mass.

The numerical results for the structure functions $V_i$, $A_i$ 
at order $p^4$ are collected in the first column of 
Tables~\ref{tab:vin} and \ref{tab:ain}, at $F^2=F_\pi F_K$.
[Like for the analysis of $\rR$ in the previous section,
we use the central values for $L_9^r$, $L_{10}^r$ as displayed in 
Appendix~\ref{app:notation} and neglect their uncertainties, which are
generously taken into account through the uncertainties that we will attach 
to higher order terms.]
The axial terms  have only tree-level
contributions at this order and are  strictly constant. As for the vector
terms, the variation in the $V_i$ indicated by the error range in the
first column of Table~\ref{tab:vin} is very small.
In fact $V_1$, $V_2$ are dominated by the counterterm contributions 
(at a typical scale like $\mu=M_\rho$),
which are necessarily constant at $\Order(p^4)$.
For comparison, we also show the numerical values for the
approximations in \eqref{Vit=0} in the column dubbed
accordingly. 

\begin{table}
\centering
\caption{Average values for the vector amplitudes $V_i$ in 
\protect{\eqref{eq:ViSD}}.
  The symbol $\av{V_i}$ denotes the average of the {real part}
  of the $V_i$ in units of $M_K$.
The first column displays the result  at order $p^4$ 
(with the size of variation over phase space indicated), and the second column 
gives the values in the approximation given in \eqref{Vit=0}. 
 The last column contains an estimate of higher order contributions, 
see main text for details.
\label{tab:vin}
}
\medskip
\renewcommand{\arraystretch}{1.4}
\begin{tabular}{crclrc}
\hline
  & \mcc{$\Order(p^4)$} & \eqref{Vit=0}& uncertainty \\
\hline
$\av{V_1}$ & $-1.26$\nl$\pm$\nl$0.004$& $-1.25$& $\pm 0.4$ \\ 
$\av{V_2}$ & $ 0.12$\nl$\pm$\nl$0.002$& $ 0.12$& $\pm 0.2$ \\
$\av{V_3}$ & $-0.02$\nl$\pm$\nl$0.001$& $-0.02$& $\pm 0.1$ \\
$\av{V_4}$ & $ 0   $\nl$   $\nl$     $& $ 0   $& $\pm 0.1$ \\
\hline 
\end{tabular}
\renewcommand{\arraystretch}{1.0}
\end{table}

\subsection{Numerical evaluation of the \boldmath{$\Order(p^6)$} results\label{sec:Np6}}

It is  desirable
to get a handle on the typical size of the corrections to be expected at 
$\Order(p^6)$, and we start the discussion with the axial terms that we have
evaluated analytically at order $p^6$.
 Our numerical estimate for these terms
is obtained by taking their real parts  at the
scale $\mu=M_\rho$, averaged over phase space as before.
 The contributions from the counterterms 
 are the essential uncertainty. 
We have estimated the order of magnitude of these polynomial terms in
the following manner.
The low-energy constants depend logarithmically 
on the renormalization scale~\cite{p6anom}.
It seems unnatural for the constants to be much smaller than
the change induced by the running of the scale, e.g.\ changing the logarithms
by one unit. The shifts in the polynomial contributions
of $A_{1,2}$ induced by a change of the logarithm by one unit is the following, 
\begin{align}
A_{1,\rm ct} &\eq \pm \frac{1}{192\pi^4F_\pi^2 F_K^2} 
 \Bigl\{ 7 M_K^2 - 7 M_\pi^2 + s + t - 2u \Bigr\}
~, \nonumber\\
A_{2,\rm ct} &\eq\pm\frac{1}{768\pi^4F_\pi^2 F_K^2} \Bigl\{ 
   25 M_K^2 - 17 M_\pi^2 - 7 t - 8 u \Bigr\}
~. 
\end{align}
In both cases, there are (potentially) large $M_K^2$
corrections that could dominate the $\Order(p^6)$ contributions.
As $A_4$ consists exclusively of a counterterm contribution that is
scale independent by itself, the above procedure cannot be applied here. 
We use instead an even rougher dimensional estimate
\beq
A_{4,\rm ct} \eq \pm \frac{16}{(4\pi)^4F_\pi^2F_K^2} ~.
\eeq 
Finally, we do an average of these polynomial terms as before
and quote the result in the second column of Table~\ref{tab:ain} as the final
uncertainty at this order. We have not worked out the amplitude $A_3$ at order
$p^6$ because it is only needed for a comparison with the amplitudes in
\cite{ktev} at order $p^4$ 
and drops out in the basis used in the present work.

In order to generate an analogous estimate of the contributions at order $p^6$
 for the vector terms, a two-loop calculation is needed. This is beyond the
scope of the present work, and we content ourselves here with the 
rough estimates displayed in the last column of Table~\ref{tab:vin}. These are
obtained as follows. Concerning $V_1$, we estimate the contributions at order
$p^6$ and higher to be of the order of 30$\%$ of the leading term.
 As $V_2$ is suppressed at leading order,
we scale its uncertainty by a factor of 2. Finally, for dimensional reasons,
the counterterm contributions to $V_{3,4}$ are  constant. The numbers
displayed in the last column for $V_{3,4}$ are obtained from a
dimensional estimate similar
to the one for $A_4$ discussed above.
\begin{table}
\centering
\caption{Average values for the axial amplitudes $A_i$ in 
  \protect{\eqref{deca}},
  as given by $\Order(p^4)$ and  $\Order(p^6)$ ChPT.
  The central values in the second column refer to the order $p^6$ 
result at the scale $\mu=M_\rho$, with the counterterms set to zero. 
 For the estimates of the uncertainties, see main text.
  The symbol $\av{A_i}$ denotes the average of the {real part}
  of  $A_i$ in units of $M_K$. 
The term $A_3$ was not determined at order $p^6$ for
  reasons explained in the text.\label{tab:ain}
}
\medskip
\renewcommand{\arraystretch}{1.4}
\begin{tabular}{crrcl}
\hline
  & $\Order(p^4)$ &  \mcc{$\Order(p^6)$} \\
\hline
$\av{A_1}$ & $ 0   $ & $ -0.07$\nl$\pm$\nl$0.2$ \\ 
$\av{A_2}$ & $-0.30$ & $ -0.25$\nl$\pm$\nl$0.1$ \\
$\av{A_3}$ & $ 0   $ & && \\
$\av{A_4}$ & $ 0   $ & $  0   $\nl$\pm$\nl$0.4$ \\
\hline 
\end{tabular}
\renewcommand{\arraystretch}{1.0}
\end{table}

\subsection{Predictions for the amplitudes\\ used in previous analyses\label{sec:ktevCD}}

In the recent analysis~\cite{ktev} of radiative $K_{e3}$ decays,
 the IB part was taken from \cite{FFS}. It differs from the 
one displayed in \eqref{eq:VIB}
through terms of order $q$ and higher, and can be obtained 
from $V^{\rm IB}_{\mu\nu}$ by subtracting  these additional
 terms. In addition, a different basis 
of transverse tensors for the SD part was used. In this
 subsection, we first discuss the relation between 
the Lorentz invariant
 structure functions in the two conventions, 
and then elaborate on their  chiral expansion.

The IB part used in~\cite{ktev} is 
\bea\label{eq:VIBFFS}
\vibf&=&\frac{p'_\mu}{p'q}
\Bigl(2p_\nu \bigl\{ f_+-2qW\dot f_+\bigr\}-W_\nu \bigl\{f_2-2qW\dot f_2 \bigr\}\Bigr)
\nonumber\\[2mm]
&+&
2W_\mu\Bigl(2p_\nu\dot f_+ - W_\nu \dot f_2\Bigr)
-g_{\mu\nu}f_2 ~\fs 
\eea
The form factors  $f_i$ as well as their derivatives 
$\dot{f}_i=df_i/dt$ are 
evaluated with argument $t$. Here and below, barred quantities indicate that the
convention from \cite{FFS},  \eqref{eq:VIBFFS}, is used for
the inner bremsstrahlung part. 
The SD part  changes accordingly, such that the sum of IB and SD
remains the same,
\bea
\vmn=\vibf+\vsdf\fs
\eea
Four of the eight Lorentz invariant amplitudes  were
retained in~\cite{FFS},\cite{ktev}  
and denoted by $A,B,C$ and $D$. Here, we extend this notation to
the remaining four amplitudes and write
\begin{align}
\vsdf &\eq 
  \frac{A}{M_K^2} \bigl(p _\mu q_\nu - p q\,g_{\mu\nu}\bigr)
+ \frac{C}{M_K^2} \left(p'_\mu q_\nu - p'q\,g_{\mu\nu}\right) \no\\
& \hskip 0.5cm
+ \left(p'q\,p_\mu - pq\,p'_\mu\right) 
  \biggl\{ \frac{E}{M_K^4}\, p'_\nu + \frac{G}{M_K^4}\, p_\nu \biggr\} ~, \no\\
A_{\mu\nu} &\eq i\,\epsilon_{\mu\nu\rho\sigma} 
  \biggl\{ \frac{B}{M_K^2}\, p^\rho + \frac{D}{M_K^2}\, p'^\rho \biggr\}
  q^\sigma \label{defabcd} \\
& \hskip 0.5cm
+ i\,\epsilon_{\nu\rho\sigma\lambda}\,p^\rho p'^\sigma \biggl\{
  \left(p'q\,g_\mu^\lambda-p'_\mu q^\lambda\right) \frac{F}{M_K^4} \no\\
& \hskip 2.7cm
+ \left(p q\,g_\mu^\lambda-p _\mu q^\lambda\right) \frac{H}{M_K^4} \biggr\} ~.\no
\end{align}
The relation to the  $V_i,A_i$ used here is
\beq \begin{split}
A &\eq M_K^2 \Bigl(  \vsdifc{2} + p'q \, \vsdifc{3}\Bigr)~,\\
B &\eq - M_K^2 \Bigl( A_2 + p'W  A_3 + M_\pi^2 A_4 \Bigr)~,\\
C &\eq  M_K^2 \Bigl( \vsdifc{1} -\vsdifc{2} - pq \, \vsdifc{3} \Bigr)~,\\
D &\eq M_K^2 \Bigl( A_1 + A_2 + pW A_3 + pp' A_4 \Bigr) ~,\\
E &\eq  M_K^4 \bigl( \vsdifc{3} - \vsdifc{4} \bigr)~,\\
F &\eq M_K^4 \bigl( A_3-A_4 \bigr)~,\\
G &\eq - M_K^4 \, \vsdifc{3}~, \\
H &\eq  - M_K^4\, A_3~, \\
\end{split}
\label{relabcd} \eeq
where
\bea
\vsdifc{1}&=&V_1-\frac{2\triangle_2 f_+}{p'q},\hspace{2.5cm}
\vsdifc{2}=V_2-4\dot f_+\scs\nnnl
\vsdifc{3}&=&V_3-\frac{2\triangle_2 f_+-\triangle_2 f_2}{p'q\,qW},\hspace{1cm}
\vsdifc{4}=V_4-\frac{2\triangle_2 f_+}{p'q\,qW}\scs\nnnl\triangle_2 f_i
&=&f_i(t)-f_i(W^2)-2qW\dot f_i\fs \label{VbartoV}
\eea 
Equation~\eqref{relabcd} displays the transformation between the 
basis used in \cite{ktev} and in the present work,
 while \eqref{VbartoV} presents the changes induced by  the difference 
in the IB part.\footnote{
In order to  check the sign conventions used here and
in~\cite{FFS},\cite{ktev} -- where the Pauli metric is used --
we have algebraically 
evaluated the expression of the decay width with 
\eqref{eq:VIBFFS}, \eqref{defabcd}
in terms of $f_+,f_2, A,B,C,D$ and 
in the limit of vanishing electron mass. We 
found complete agreement with the corresponding expressions given in 
(A1)--(A5) of \cite{FFS}, up to an obvious 
 misprint in the  line after (A3). 
The amplitudes $E$--$H$ were not used in~\cite{FFS,ktev}.}

The above relations allow us to calculate the phase space averaged structure
functions $\av{A},\,\av{B},\,\ldots$ in a \linebreak  straightforward
manner.  To be specific, we use
linear form factors $f_+,\,f_2$, as a result of which only the derivative term
$\dot f_+$ in $\vsdifc{2}$ matters. The final  
result is displayed in
Table~\ref{tab:abcd} where, for reasons that become 
clear at the end of this subsection, we stick to the
values at order $p^4$ in the chiral expansion. In this approximation, the
axial terms are constant -- this is why we do not display an error band in the
last column in Table~\ref{tab:abcd}.
\begin{table}
\centering
\caption{Values of the structure-dependent terms in the KTeV
  conventions, as given by $\Order(p^4)$ ChPT [with the size of variation over
  phase space indicated].\label{tab:abcd}
}
\medskip
\renewcommand{\arraystretch}{1.4}
\begin{tabular}{crclcrcl}
\hline
$\av{A  }$ & $-1.34$\nl$\pm$\nl$0.002$&
$\av{B  }$ & $ 0.30$\nl$   $\nl$     $\\
$\av{C  }$ & $ 0.08$\nl$\pm$\nl$0.005$&
$\av{D  }$ & $-0.30$\nl$   $\nl$     $\\
$\av{E  }$ & $-0.02$\nl$\pm$\nl$0.001$&
$\av{F  }$ & $ 0    $\nl$   $\nl$     $\\
$\av{  G}$ & $ 0.02$\nl$\pm$\nl$0.001$&
$\av{  H}$ & $ 0    $\nl$   $\nl$     $\\
\hline
\end{tabular}
\renewcommand{\arraystretch}{1.0}
\end{table}
We now comment on some basic features of the 
choice \eqref{eq:VIBFFS} for the IB part and \eqref{defabcd}  for 
the transverse tensors.
 As for the impact of the difference in the  IB part, we note that, 
 expanding in \eqref{relabcd} the form factors $f_i(W^2)$ 
around $q=0$,
 it is readily seen that  the SD amplitude $\vsdf$ indeed 
differs from the one in the
present work only by terms of order $q$ and higher,
 as it must be for a reasonable choice of IB.
 On the other hand, as already mentioned in Subsect.~\ref{sec:ibsd}, 
these additional terms are singular at $s=M_\pi^2$, and can
potentially distort the amplitudes near the boundary of phase space.
The difference in the choice of the SD part 
 generates more  pronounced effects.
 As we have mentioned before,  the 
structure functions $V_3$, $A_3$ are suppressed by a factor
of $m_e^2/M_K^2$ and are therefore inaccessible in the electron decay mode.
The tensor decomposition \eqref{defabcd} does not make use of this 
fact.
 As a result,  certain simultaneous shifts in $A$, $C$, $E$, $G$
(corresponding to a change in $V_3$) or simultaneous shifts in $B$, $D$, $F$,
$H$ (corresponding to a change in $A_3$) are unobservable.
Measurable combinations are
\bea
A + \frac{p'q}{M_K^2}\,G &\eq&   M_K^2 \, \vsdifc{2} ~, \no\\
B - \frac{p'W}{M_K^2}\,H &\eq& - M_K^2 \bigl( A_2 + M_\pi^2 A_4 \bigr) ~, \no\\
C - \frac{p q}{M_K^2}\,G &\eq&   M_K^2 \bigl( \vsdifc{1} - \vsdifc{2} \bigr) ~, 
\label{eq:abcdphys} \\
D + \frac{p W}{M_K^2}\,H &\eq&   M_K^2 \bigl( A_1 + A_2 + pp' A_4 \bigr) ~, \no\\
E + G &\eq& - M_K^4 \, \vsdifc{4} ~, \no\\
F + H &\eq& - M_K^4 \, A_4 ~. \no
\eea
In other words, the decay width can be expressed in terms of the quantities on
the left hand side of \eqref{eq:abcdphys}.
We conclude that e.g.\ the
structure functions $C,D$ -- or any linear combination thereof -- 
are not measurable in $K_{e3\gamma}$, as long as $V_3,A_3$ are nonzero.
  In the following
section, we discuss this point in some more detail. 
In particular, we will provide an  interpretation of
the quantities $C$ and $D$ determined by the KTeV Collaboration in 
\cite{ktev}.

Finally, coming back to Table~\ref{tab:abcd}, we note that, 
because $A,B,C$, and $D$ are not observables, it does 
not make much sense to work out their numerical magnitude at order $p^6$.
On the other hand, their value at order $p^4$ will be of use in 
the following section.

\section{Structure-dependent terms in differential rates}
\label{sec:distributions}

\subsection{\boldmath{$\Eg$} distribution: theory\label{sec:Egdist}}
Of the various differential rates one may consider, the distribution
$d\Gamma/d\Eg$  stands out for the purpose 
of extracting information on structure-dependent terms,
as $\Eg$
is the very variable to distinguish bremsstrahlung and the 
structure-dependent part of the amplitude.  
In our investigation, we shall neglect the terms coming
from the square of the structure-dependent amplitude $T^{\rm SD}$.
Furthermore, we make use of the observation made in the previous
section that in the one-loop approximation, 
these structure functions 
are constant to rather high accuracy: we replace them 
in the expression \eqref{eq:traces} for the square of the matrix element
by the averages $\av{V_i},\av{A_i}$. We then obtain 
 the following decomposition of the photon spectrum:
\beq \begin{split}
\frac{d\Gamma}{d\Eg} = \frac{d\Gamma_{\rm IB}}{d\Eg}
&+ \sum_{i=1}^4 \left( \av{V_i} \,\frac{d\Gamma_{V_i}}{d\Eg}
                    + \av{A_i} \,\frac{d\Gamma_{A_i}}{d\Eg} \right) \\
&+\Order\Bigl(|T^{\rm SD}|^2,\,\Delta V_i,\,\Delta A_i\Bigr) \fs
\end{split} \label{specdecomp}
\eeq
The quantity  ${d\Gamma_{V_i}}/{d\Eg}$ denotes the part 
of the spectrum
 which is proportional to $\av{V_i}$,  
and analogously for \linebreak
${d\Gamma_{A_i}}/{d\Eg}$.
[Remember that we define $\av{V_i}$, $\av{A_i}$ to be dimensionless.]
 The quantities $\Delta V_i,\Delta A_i$ stand for the errors
introduced by this  approximation. 

In the following, we shall neglect the effect of $V_4$ 
and $A_4$.\footnote{We have verified that the
  distributions for $V_4$ and $A_4$ are indeed considerably smaller than the
  ones discussed here, in addition to the fact that both $\av{V_4}$
  and $\av{A_4}$ vanish at leading chiral order.  This holds for all
  differential rates discussed here 
and in Sect.~\ref{sec:otherdist}.}
The objective is to study the distributions
$d\Gamma_{V_i}/d\Eg$, $d\Gamma_{A_i}/d\Eg$ in order to
quantify the possibility to extract  $\av{V_i}$ and 
$\av{A_i}$ from data. In order to obtain experimental information
independent of the measurement of the total rate, we follow the
strategy of~\cite{ktev} and only discuss spectra with arbitrary
normalization.  
Furthermore, we follow the procedure in that publication and
deviate here from the ``standard cuts'', instead we use
$\tecut=5^\circ$.
We have found, though, that such a reduction of the angle cut only
increases the size (and therefore the expected statistics in an
experiment) of the bremsstrahlung and hardly has any effect on the
structure-dependent spectra.
\begin{figure}
\vskip 2mm
\centering
\includegraphics[height=2.8in]{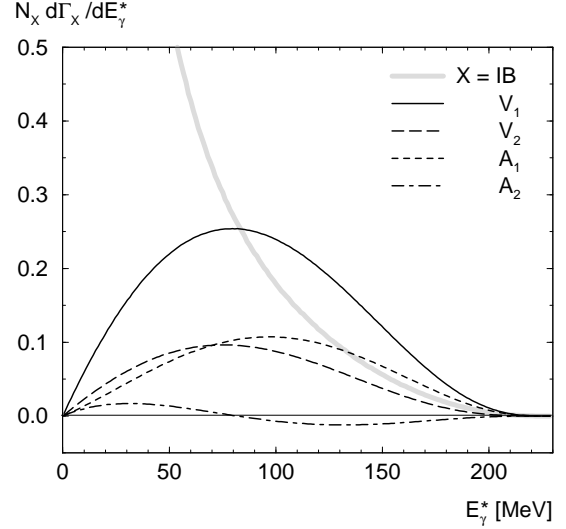} 
\caption{Photon energy distributions from inner bremsstrahlung
  as well as the various structure-dependent terms.  The notation
  $d\Gamma_X/d\Eg$ for the various $X$ refers to
  \eqref{specdecomp}.  The normalization factors are
  $N_{V_i,\,A_i}=200\,N_{\rm IB}=10^3M_K/\Gamma(K_{e3})$. 
  We only cut on the electron-photon angle,
  $\tecut=5^\circ$~\cite{ktev}. \label{fig:Egdist}}   
\end{figure}
The relevant photon spectra are shown in Fig.~\ref{fig:Egdist}.  
Note that the bremsstrahlung distribution is scaled down by a factor  
of 200 relative to the structure-dependent parts. 
We observe the expected fall-off of $d\Gamma_{\rm IB}/d\Eg
\propto 1/\Eg$ as well as the linear rise of all 
structure-dependent spectra for small photon energies.  As phase space bends
them down to zero at maximum photon energy, all structure-dependent
distributions show a maximum (a maximum and a minimum in the case of
$A_2$), which for $V_1$, $V_2$ occurs around $\Eg = 80$~MeV, for
$A_1$ slightly higher, around $\Eg = 100$~MeV.  Although the
$A_2$ spectrum has a form distinct from all others, its magnitude is
far too small to be observable.  In view of the chiral $\Order(p^4)$
prediction $A_1=0$, this means that no effects of the chiral anomaly
are likely to be extracted from the photon energy spectrum.

The remaining three structure-dependent spectra are remarkably similar
in shape, if not in height.  If we assume that the experimental
accuracy is not sufficient to observe the slightly shifted positions
of the maxima in the three spectra, we have approximately
\beq\label{eq:fegamma}
f(\Eg)
\;\doteq\;            \frac{d\Gamma_{V_1}}{d\Eg} 
\;\approx\; 2.6 \times\frac{d\Gamma_{V_2}}{d\Eg} 
\;\approx\; 2.4 \times\frac{d\Gamma_{A_1}}{d\Eg}  
~,
\eeq
where we have taken the height of the peaks as the measure for the
proportionality factors, irrespective of the exact energy where they
occur.
[In case that more accurate data is available, it would be straightforward
to incorporate a more refined representation of the photon spectrum than the
one proposed here.]

Equation~\eqref{eq:fegamma}  is the main result of our
investigation of the photon spectrum:
\begin{enumerate}
\item To good approximation, the photon energy spectrum originating from the
  bremsstrahlung amplitude is distorted 
by \emph{one single function} $f(\Eg)$.
 The information on the SD terms is contained 
  in the effective strength  $\av{X}$ that 
multiplies $f(\Eg)$,
\beq \begin{split}
\frac{d\Gamma}{d\Eg} &\,\approx\,
\frac{d\Gamma_{\rm IB}}{d\Eg}  + \av{X} \;f(\Eg)  ~, \\
\av{X} &\eq     \av{V_1} 
        + 0.4 \,\av{V_2}
        + 0.4 \,\av{A_1} \label{Vicombine} ~.
\end{split} \eeq
\item 
The three amplitudes
 $V_1$, $V_2$, $A_1$ 
differ mainly in terms of the
  \emph{weight} with which they contribute to $\av{X}\fs$
  The latter can be calculated in ChPT,
\bea\label{eq:Xnum}
 \av{X} \eq
\left\{ \begin{array}{ll}
 -1.2         & O(p^4) \\[2mm]
 -1.2 \pm 0.4 & O(p^6) ~.
\end{array} \right.
\eea
Note that the uncertainty in the contribution from the vector channel 
has only been roughly estimated here.
\item\label{it:measure}
In order to \emph {measure} 
$\av{X}$, one may  use the representation \eqref{Vicombine} for the
spectrum, insert the explicit form of $f(\Eg)$
and do a fit to the data with $\av{X}$ as a free parameter.
Alternatively, one may take any of the amplitudes $V_{1,2}$  or $A_1$, take 
it to be  constant over phase space, and perform
 a fit to the bremsstrahlung spectrum. The result
will be the same. However, it is clear that in this manner, one has not
determined the chosen amplitude to  perform the fit,
 but just the effective strength $\av{X}$.
\end{enumerate}

\subsection{\boldmath{$\Eg$} distribution: experiment\label{sec:ktevdist}}

We now discuss the
result of the KTeV analysis~\cite{ktev} in light of the 
previous subsection.
First, we note that in~\cite{ktev}, all SD
parts were set to zero, except the amplitudes $C,D$, 
that were taken to be constant over phase space. This amounts to the procedure
mentioned in point~\ref{it:measure}.\ above, except that two amplitudes
have been retained in~\cite{ktev}, while one is sufficient to measure $\av{X}$.
Indeed, \cite{ktev} finds a strongly eccentric
error ellipse constraining the parameter
space for these two structure-dependent terms.
In order to compare the KTeV result with the above 
representation of $\av{X}$,
we translate the KTeV amplitudes into our conventions.
We assume
a linear form factor $f_+$, use the relations \eqref{relabcd}
and find that, with $A=B=0\,$,
\beq\label{eq:translatecd}
V_1 \eq C/M_K^2 ~,~~ V_2\eq 4\dot f_+(0) ~,~~ A_1\eq D/M_K^2 ~.
\eeq
The $V_i,A_i$ not listed are zero. In other words, the amplitudes 
\eqref{eq:translatecd} 
result in the
same photon spectrum as the one generated by the amplitudes
used in~\cite{ktev}. We therefore conclude that the effective strength
$\av{X}$ is given in this case by
\bea\label{eq:Xktev}
\av{X}\eq {C}+\tan{(23^\circ)} {D}+1.5M_K^2\dot f_+(0)\scs
\eea
where we have dropped the bracket notation  for  $C,D$, because
$\av{C}=C$ for constant amplitudes, and the angle is introduced for easy
comparison with \cite{ktev}.
 The structure of \eqref{eq:Xktev} 
has been  confirmed  by the observation made 
in~\cite{ktev} that it is
\bea\label{eq:linktev}
C'=\cos{(25.8^\circ)} \bigl[ C +\tan{(25.8^\circ)}D \bigr] 
\eea
which is best constrained by the 
data, with~\cite{ktev}
\bea\label{eq:valktev}
C'&=&-2.5^{\,\,+1.5}_{\,\,-1.0}(\text{stat}) 
\pm 1.5 (\text{syst}) ~.
\eea
This may be compared with the calculation in the framework of ChPT. Using 
\eqref{eq:Xktev} and \eqref{eq:Xnum}, and neglecting the small difference
in the angle, we find
\bea
C'=-1.6\pm 0.4\qquad{\mbox{\small{[ChPT]}}}\scs
\eea
which agrees with \eqref{eq:valktev} rather well.

While an interpretation of the KTeV result \eqref{eq:valktev} 
as a measurement of the effective
coupling is sound, it does not allow one to draw conclusions about the size of
the SD terms themselves because, as we have shown in the previous section, $C$
and $D$ are not observable amplitudes as long as the amplitude $V_3$ is not
negligible. 
 In addition, the assumption $A=B=0$ made in the analysis
of~\cite{ktev}, on the basis of the {\it soft kaon approximation}, 
is incorrect and invalidates such an interpretation of $C'$ even for a
negligible $V_3$. 
Chiral perturbation theory may be used to illustrate this point:
we consider the amplitudes at order $p^4$ and
disregard the structure function $V_3$ altogether, then
from \eqref{relabcd}, we find
\beq \begin{split}
\av{X} &\eq \underbrace{{1.4\,\av{A}}}_{-1.9}
+\underbrace{{0.4\,\av{B}}}_{+0.1}
+\underbrace{{     \av{C}}}_{+0.1}
+\underbrace{{0.4\,\av{D}}}_{-0.1}
+\underbrace{{1.5\,\mk \,\dot f_+(0)}}_{+0.6} \\
&\eq-1.2 ~,~
\end{split}\label{eq:softkaon}
\eeq
where we have again used the phase space average for the structure functions,
because they are not constant in this case.
Equation~\eqref{eq:softkaon} shows that the main contribution to the effective 
strength $\av{X}$ is due to the amplitude $A$, while ${C}$ 
plays a minor role, and the contribution from $D$ is canceled by the one from 
$B$.  Therefore, the approximation of setting $A$ and $B$ to zero is not
valid and, consequently, $\av{X}$ is not
dominated by $\av{C}+0.4\av{D}$, and should not be taken as a
measure of this combination of amplitudes.

\subsection{Systematic errors}

\begin{figure}
\vskip 2mm
\centering
\includegraphics[height=2.8in]{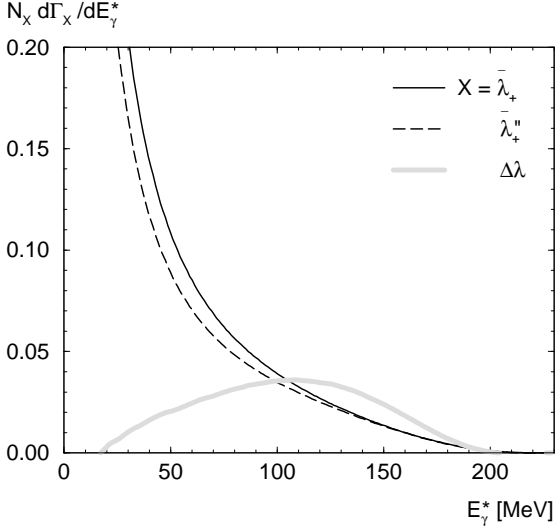} 
\caption{Photon energy distributions from inner bremsstrahlung,
  proportional to $(\bar{\lambda}_+-1)$ and $\bar{\lambda}''_+$, 
  as well as a fine-tuned difference between the two.
  The normalization factors are given by
  $N_{\Delta\lambda}=100\,N_{\bar{\lambda}_+}=10\,N_{\bar{\lambda}''_+}
  =10^3M_K/\Gamma(K_{e3})$. 
  We cut on the electron-photon angle with $\tecut=5^\circ$~\cite{ktev}.
  \label{fig:Egffback}}  
\end{figure}

We now discuss one potential source for systematic errors in this
procedure of determining the effective strength $\av{X}$ and start with the
observation that the analysis obviously requires a rather precise knowledge
of $d\Gamma_{\rm IB}/d\Eg$.  As we consider unnormalized
spectra, we are insensitive to overall coupling constants and
$f_+(0)$, but we should investigate whether a shift in the 
$K_{e3}$ form factor parameters $\lambda_+$, $\lambda''_+$ can
simulate a contribution to the spectrum similar to the 
structure-dependent effects.  For this purpose, we expand a general
bremsstrahlung spectrum with arbitrary form factor around
our choice for these parameters,
\beq
\frac{d\Gamma_{\rm IB}}{d\Eg} \eq
\frac{d\Gamma_{\rm IB}}{d\Eg}
\Biggr|_{\!\!
         \begin{array}{l}\scriptstyle\bar{\lambda}_+  =1 \\[-1mm]
                         \scriptstyle\bar{\lambda}''_+=0 \end{array}}
\!+\, \bigl(\bar\lambda_+ -1 \bigr) 
   \, \frac{d\Gamma_{\bar{\lambda}_+}}{d\Eg} 
\,+\, \bar \lambda''_+
   \, \frac{d\Gamma_{\bar{\lambda}''_+}}{d\Eg} 
\,+\, \ldots ~.
\label{ffspectra}
\eeq 
The two spectra 
${d\Gamma_{\bar{\lambda}_+}}/{d\Eg}$ and 
${d\Gamma_{\bar{\lambda}''_+}}/{d\Eg}$
are displayed in Fig.~\ref{fig:Egffback} with a solid and a dashed line, respectively.
We observe that they are
rather sizeable, but very similar in shape to the overall IB
spectrum.  Only a fine-tuned cancellation of both can lead to a
peak-like structure, which we find for
\beq
\alpha \,\doteq\, \frac{1-\bar\lambda_+}{\bar\lambda''_+} 
\eq 0.080 \pm 0.005  ~.\label{dangerous}
\eeq
Stated differently, $\alpha$ has to be
within this narrow range  in order for the combination
\beq\label{eq:peaked}
\frac{d\Gamma_{\Delta\lambda}}{d\Eg} \;\eq\;
\frac{d\Gamma_{\bar{\lambda}''_+}}{d\Eg}
- \alpha \, \frac{d\Gamma_{\bar{\lambda}_+}}{d\Eg} 
\eeq
 to have
a maximum. 
  This means that, for example, a simultaneous reduction of $\lambda_+$ by
8\% as compared to the central input value and the introduction of a
quadratic term in the form factor with $\lambda''_+=(\lambda_+^c)^2$ 
(as suggested by $K^*$ pole saturation) mimics a structure-dependent
contribution with a peak at roughly similar energies as $f(\Eg)$.

The distribution (\ref{eq:peaked})
 is displayed  in Fig.~\ref{fig:Egffback} as a grey band at $\alpha=0.08$.
Note that  the strength of the peak is not big:  for the chosen
combination, it is about 10\% of the dominant spectrum
$\av{V_1}d\Gamma_{V_1}/d\Eg$.  
 To illustrate potential effects of this ``background'', 
we compare these findings to the latest KTeV form factor
measurements~\cite{ktevlambda}, $\bar{\lambda}_+ = 0.70 \pm
0.06$, $\bar{\lambda}''_+ = 1.85\pm 0.40$.\footnote{Note the 
conflicting results in~\cite{Lai}.}
 Taking into account the correlation~\cite{ktevlambda} 
 between $\bar{\lambda}_+$
 and $\bar{\lambda}''_+$, we find that they lead to
$\alpha = 0.16 \pm 0.01$, and we conclude the following:
\begin{enumerate}
\item
  Although these values for $\bar\lambda_+,\,\bar\lambda''_+$ are very
  different from our assumed central ones, they do not lead to a
  peak-like structure.
\item
  Even in the worst possible case with $\alpha \approx 0.08$ and
  $\bar\lambda''_+ \approx 2$, the value for $\av{X}$ based on
  the assumptions $\bar\lambda_+=1,\,\bar\lambda''_+=0$   is less
  negative than the true one. In other words, the modulus of $\av{X}$ would be
  even \emph{bigger} in the real world, by (20-25)\%.
\end{enumerate}
A more detailed analysis of this background phenomenon
ought to be performed on real data.

\subsection{Other distributions\label{sec:otherdist}}

We have emphasized that the study of the photon energy
spectrum, at least with the currently achievable statistics, seems to give
access to only one specific linear combination of structure-dependent terms,
which is most sensitive to $V_1$.
Ideally one would find alternative distributions that are more sensitive to
the other terms $V_2$, $A_1$, $A_2$ in order to achieve a complete
decomposition into the four (main) structure functions. 
The strategy for studying the various possible differential rates is to find
those structure-dependent contributions that differ in shape from inner
bremsstrahlung and, in contrast to $d\Gamma/d\Eg$, from each other.  
We have studied differential rates with respect to the other four independent
variables $\Ep$, $\Ee$, $x$, $W^2$, but also to related variables $s$, $t$,
$u$, $\cos\te$, $\cos\tp$. 
Where applicable, we have used the cuts $\Ecut=25$~MeV, 
$\tecut=5^\circ$ in analogy to the procedure
in~\cite{ktev}.

One general feature can already be seen from the \linebreak $d\Gamma/d\Eg$ plots and
appears in almost all distributions:   the relative importance of $V_1$, $V_2$,
$A_1$, $A_2$ is the same in most cases, as the integral over a differential rate
has to be the same, no matter what kinematical variable is studied.  Therefore
distributions tend to be most sensitive to $V_1$, followed by $V_2$ and $A_1$
at roughly equal strength.  The exception to this rule is $A_2$ that shows a
sign change in most distributions, but again in most cases it is suppressed
with respect to the other structure-dependent terms by at least one order of
magnitude.

We shall only discuss those differential rates in some detail that seem to have
interesting features.  The distributions in $W^2$, $\cos\tp$
seem to offer no promising possibilities to extract information on any of the 
structure-dependent terms as their distributions are too similar to
the dominant bremsstrahlung one, while those in $\Ee$, $x$, or $s$ show $A_2$ curves
that have interesting shapes (usually with an additional zero), but are
probably far too much suppressed.

\begin{figure}
\vskip 2mm
\centering 
\includegraphics[height=2.8in]{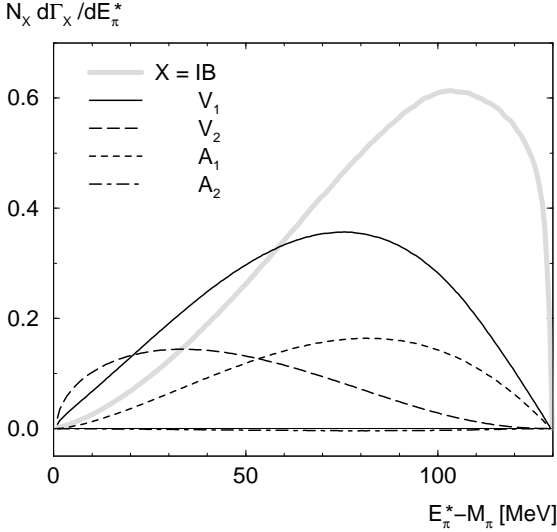} 
\caption{Pion energy distributions from inner bremsstrahlung
  as well as the various structure-dependent terms.  The notation
  $d\Gamma_X/d\Ep$ is chosen in analogy to 
  \eqref{specdecomp}.  The normalization factors are
  $N_{V_i,\,A_i}=200\,N_{\rm IB}=10^3M_K/\Gamma(K_{e3})$. 
  The cuts $\Ecut=25$~MeV, 
  $\tecut=5^\circ$ were applied.
  \label{fig:Epdist}}  
\end{figure}
More promising seem to be the partial rates $d\Gamma/d\Ep$ that are
displayed in Fig.~\ref{fig:Epdist}.  There is no divergent behavior
visible in these distributions, all of them vanish at minimum and
maximum pion energies, and the partial rates for bremsstrahlung as
well as for $V_1$, $V_2$, and $A_1$ have one peak in the spectrum 
($A_2$ is nearly completely suppressed here).  
We observe that the bremsstrahlung distribution is peaked at high pion
energies (for $\Ep-M_\pi \approx 100$~MeV), and 
so are the $V_1$ and $A_1$ partial rates, even though their
respective peaks occur a bit lower.  
Distinct from all these is, however, the $V_2$ contribution to the
pion energy distribution that is peaked at \emph{small} pion energies. 
Although the overall sensitivity is
again by roughly a factor of 3 smaller than that for $V_1$, this
partial rate might be a window to access information on the structure
function $V_2$.

We remark, though, that this extraction might again be obscured
by uncertainties in the form factor $f_+$:  
the distributions $d\Gamma_{\bar{\lambda}_+}/d\Ep$ and
$d\Gamma_{\bar{\lambda}''_+}/d\Ep$, defined in complete analogy to
\eqref{ffspectra}, also turn out to be peaked for
lower pion energies than the total bremsstrahlung distribution.
Of course this problem can be remedied by more precise form factor data as
provided e.g.\ in~\cite{ktevlambda,Lai}.   
A more detailed study should be performed with actual experimental data.

\begin{figure}
\vskip 2mm
\centering 
\includegraphics[height=2.8in]{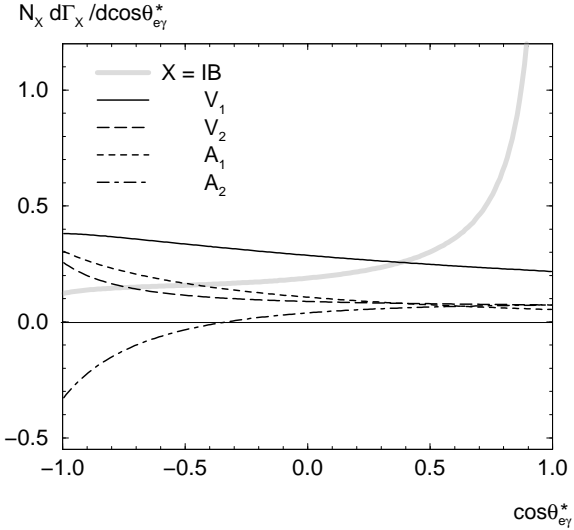} 
\caption{Distributions with respect to $\cos\te$ from
  inner bremsstrahlung
  as well as the various structure-dependent terms.  The notation
  $d\Gamma_X/d\cos\te$ is chosen in
  analogy to \eqref{specdecomp}.  The normalization factors are
  $N_{V_i,\,A_i}=200\,N_{\rm IB}=10^4/\Gamma(K_{e3})$. 
  The photon energy cut $\Ecut=25$~MeV was applied.
  \label{fig:CTegdist}}  
\end{figure}
There is special interest in finding a partial rate with a more
pronounced contribution from $A_2$ for the following reason:  as
discussed in Subsect.~\ref{sec:Np4}, this is the only non-vanishing
contribution of the WZW-anomaly term, while $A_1$ vanishes at
$\Order(p^4)$.  We have commented before on the sign change in the
$A_2$ distributions that often leads to cancellations.
Fig.~\ref{fig:CTegdist} shows a partial rate in which $A_2$ is relatively
prominent:  its contribution becomes relatively strong in
$d\Gamma/d\cos\te$ in backward direction. 
The slope of the total structure-dependent distribution in backward direction,
which can be thought of as the \emph{second} derivative with respect to
$\cos\te$ at $\cos\te=-1$, is potentially dominated 
by $A_2$.
It seems therefore that if effects of the chiral anomaly should be
visible at all, it might be accessible in the distribution with
respect to the electron-photon angle, in backward direction.

To conclude this section, we emphasize that this study of possible additional
partial rates is by no means meant to be exhaustive.  In particular,
certain effects may only be visible in double differential rates etc.  
We defer any such more extensive study until experiments give hints
about the statistical feasibility of these various suggestions.

\section{Conclusions and outlook\label{sec:conclusions}}

In this paper, we have analyzed various aspects of $K_{e3\gamma}$ decays. 
\begin{enumerate}
\item 
In the absence of radiative corrections, 
 the decay amplitude  may be decomposed into an inner
brems- \linebreak strahlung part (IB) and a structure-dependent part (SD).
Our construction of the bremsstrahlung 
amplitude guarantees that the SD part is 
regular in the Mandelstam plane, aside from the 
branch points required by unitarity.
Structure-dependent contributions can be parametrized in 
terms of eight structure functions $V_i\scs A_i\scs i=1,4$.

\item We evaluate the expression for the width with massless spinors. In other
 words, the electron mass is set to zero in the numerator of the 
relevant terms. In this approximation, the contribution from the IB part
 can be written entirely in terms of the $K_{e3}$ form factor
 $f_+$. Furthermore, the structure functions $V_3$ and  $A_3$
 cancel out. 
\item
If this IB--SD separation is applied to the chiral $\Order(p^4)$
representation of the $K_{e3\gamma}^0$ decay amplitude provided 
in~\cite{BEG}, one obtains leading-order 
chiral predictions for the structure functions 
(they vanish at order $p^2$).  The axial terms
are constant and given in terms of the WZW anomaly, while the vector terms
receive contributions both from loop graphs and the low-energy
constants ${L}_9^r$ and ${L}_{10}^r$ of the chiral 
Lagrangian at order $p^4$~\cite{GL}.  At this order, all cuts in
the loop functions lie far outside the physical region, such that also the
vector terms can be approximated to good accuracy by constants.
\item
In order to obtain control of higher order corrections, we have analyzed
$\Order(p^6)$ contributions to the structure-dependent terms.  We have
performed a complete  calculation for the axial terms. For the vector ones,
 we have determined the $L_i \times L_j$ contributions at order $p^6$, and 
 provided a very rough estimate of the remaining diagrams.

At this order, cuts appear in the physical region, both
in the axial and in the vector  structure functions.  The corresponding
imaginary parts generate $T-$odd
contributions in some of the  decay distributions. On the other hand,
 in the cases considered in this work, they drop out.

The effect of the cuts on the real parts is  diminished by the
fact that they appear as $P$-wave rescattering (axial) or only in a tiny
corner of phase space (vector).  Dominant uncertainties arise from 
$\mk$ corrections, generated by counterterms at $\Order(p^6)$.
\item 
The most precise and stable theoretical prediction can be given 
for the ratio $\rR$
of the radiative $K_{e3}$ decay width relative to the non-radiative one.
This ratio \linebreak turns out to be very insensitive to the details of the form
factor $f_+$, such that the purely hadronic result is very precise.
Structure-dependent terms yield only a 1\% correction to the bremsstrahlung,
such that even a sizeable uncertainty in the former affects the 
precision of the total value only at the few permille level.
Our prediction for $\Ecut=30\,$MeV,  $\tecut=20^\circ$,
\bea
 R= (0.96 \pm 0.01)\times 10^{-2} ~,
\eea
deviates from the KTeV results~\cite{ktev_04,ktev},  
but agrees well with the recent measurement 
of the NA48 Collaboration~\cite{na48}, see Table~\ref{tab:experiments}.
\item 
We have investigated the possibility to measure SD terms. 
We find that  the bremsstrahlung spectrum is modified by the SD terms 
essentially by one
single function $f(\Eg)$,
 and that the different structure 
functions contribute with different strength 
to the effective coupling multiplying $f(\Eg)$.
 The KTeV analysis~\cite{ktev} confirms this observation. In their language,
 the effective coupling is obtained from the combination
\bea
C'=\cos{(25.8^\circ)}C +\sin{(25.8^\circ)}D 
\eea
of amplitudes $C,D$, with~\cite{ktev}
\bea\label{eq:ktevconcl}
C'&=&-2.5^{\,\,+1.5}_{\,\,-1.0}(\text{stat}) 
\pm 1.5 (\text{syst}) ~.
\eea
The calculation in the framework of ChPT gives
\bea
C'=-1.6\pm 0.4\qquad{\mbox{\small{[ChPT]}}}\fs
\eea
We have shown why the  result \eqref{eq:ktevconcl}
should not be interpreted as  a measurement of the amplitudes $C,D$, but
rather as a measurement of the effective coupling of the SD terms to $f(\Eg)$.
\item 
We have discussed alternative distributions over phase space in a 
qualitative manner.  In order to distinguish the vector functions $V_1$,
$V_2$, the distribution in pion energies might be used.  Effects of the chiral
anomaly are highly suppressed in most distributions. It might at best be
accessible in the differential rate with respect to the electron-photon angle
in backward direction.
\end{enumerate}
Most extensions of this work will depend on the further interplay between
experimental accuracy and theoretical desirability:  for example,
a complete calculation of the radiative corrections would be desirable.  
We have refrained here from comparing to the latest KTeV results on
$\rR$ without cuts on the photon--electron angle~\cite{ktev_04}, as
this also would necessitate special care concerning radiative corrections;
this will be considered elsewhere.
As indicated
in the above, a more detailed study of how to disentangle the various
structure-dependent terms would be possible once the extraction 
of such terms 
from experiment becomes feasible.  

The most imminent extension of this work, however, is to provide
predictions also for the other $\kl3g$ channels.  An analogous study of the charged
channel $K_{e3\gamma}^+$ is most straightforward and will be performed in due
course.
The muon channels might in principle lend easier access to structure-dependent
contributions.
The KTeV  experimental determinations of the ratio $\rR$ for
$K_L\to \pi^\pm \mu^\mp \nu_\mu \gamma$~\cite{ktev_04}, with 
 improved uncertainty with respect to the  
previous NA48 results~\cite{bender}, 
can already be considered as an interesting starting point for a more 
comprehensive study of radiative $K_{\mu3}$ decays.

\subsection*{Acknowledgements}

We thank Fabrizio Scuri for collaboration in an early stage of this work.
We enjoyed informative discussions and/or e-mail exchanges with  Douglas
R.\ Bergman, Yury Bystritsky, Eduard Kuraev, Rick Kessler,
Leandar Litov, Helmut Neufeld, Massimo Passera, Hannes Pichl, 
Martin Schmid,  Stoyan Stoynev, and Rainer Wanke. 
In particular, we \linebreak thank Alberto Sirlin for  correspondence concerning the KLN
theorem, and Konrad Kleinknecht for useful comments on the
manuscript.
This work was  supported  by the Swiss
National Science Foundation, by RTN, BBW-Contract No. 01.0357,
and EC-Contract  HPRN--CT2002--00311 (EURI\-DICE).
NP was partially supported by funds of MIUR (Italian Ministry of
University and Research) and of the Trieste University.


\renewcommand{\thefigure}{\thesection.\arabic{figure}}
\renewcommand{\thetable}{\thesection.\arabic{table}}

\begin{appendix}

\setcounter{figure}{0}
\setcounter{table}{0}
\section{Notation}\label{app:notation}
We denote the charged pion and
neutral kaon masses with $M_\pi$ and $M_K$,
respectively.
In numerical evaluations, we use
\begin{align}
M_K &\eq 497.67~\text{MeV} ~,~~
M_{\pi} \eq 139.57~\text{MeV} ~,~~ \\
m_e &\eq 0.511~\text{MeV} ~,~~
F_\pi \eq 92.4~\text{MeV} ~,~~ 
F_K \eq 1.22\,F_\pi ~. \no
\end{align} 
The $K_{e3}$ form factor is parametrized by
\beq
f_+(t) \eq f_+(0)\left[1+\lambda_+\frac{t}{M_\pi^2}
+\lambda''_+\frac{t^2}{M_\pi^4}+\cdots\right] ~.
\eeq
As explained in the main text, the precise values 
of $f_+(0)$ and  $\lambda_+$ do not matter in the present context.
For numerical evaluations, we use the parameter-free 
one-loop result~\cite{leutroos}
\beq
f_+(0) \eq 0.977 ~,
\eeq
and a central value $\lambda_+^c=0.0294$.
For the low-energy constants we take
\beq
L_9^r(M_\rho) \eq 6.3\times 10^{-3} ~,~~ 
L_{10}^r(M_\rho)= -4.9\times 10^{-3} ~.
\eeq
$L_9^r$ was chosen such that the chiral one-loop representation for
$\lambda_+$ reproduces $\lambda_+^c = 0.0294$.
The sum $L_9^r+L_{10}^r$ is then fixed from $\pi_{e2\gamma}$ decays.
We express the low-energy constants in the following, scale-independent
form~\cite{BEG}: 
\beq\begin{split}
\bar{L}_9    &\eq L^r_9   (\mu) - \frac{1}{512\pi^2}\log\frac{\mpi M_K^4\me}{\mu^8}~, \\
\bar{L}_{10} &\eq L^r_{10}(\mu) + \frac{1}{512\pi^2}\log\frac{\mpi M_K^4\me}{\mu^8}~. 
\end{split}\eeq
Again, the precise values of $L^r_9$ and  $L^r_{10}$ do not matter.

\setcounter{figure}{0}
\setcounter{table}{0}
\section{Traces and decay widths}\label{app:traces}

Here, we give the explicit expression for the sum over spins in
$|T|^2$ in the limit where
the relevant traces are evaluated at $m_e=0$, and comment on the relation
between $K_{e3\gamma}$ and $K_{e3\gamma}^0$ decays in the presence of 
$T$-odd terms.

\subsection{Traces}

We write
\beq \begin{split}
N^{-1}\sum_{\rm spins}|T|^2 \
&\eq
a_1f_+(t)^2+a_2f_+(t) \delta f_+ +a_3\delta f_+ ^{\, 2}     
\\ & \,+\,
\sum_{i=1}^4\Bigl[\bigl(b_i  \,\text{Re} V_i
                       +b^5_i\,\text{Re} A_i\bigr)f_+(t) 
\\ & \hskip 1cm +
                  \bigl(c_i  \,\text{Re} V_i
                       +c^5_i\,\text{Re} A_i\bigr)\delta f_+\Bigr] 
\\ & \,+\, \xi\,
\sum_{i=1}^4\Bigl[\bigl(d_i  \,\text{Im} V_i
                       +d^5_i\,\text{Im} A_i\bigr)f_+(t) 
\\ & \hskip 1cm +
                  \bigl(e_i  \,\text{Im} V_i
                       +e^5_i\,\text{Im} A_i\bigr)\delta f_+\Bigr] 
\\ & \,+\,
{\cal O}(V_i^2,A_i^2,V_i A_i) \scs
\end{split}\label{eq:traces} 
\eeq
with
\beq \begin{split}
\xi &\eq M_K^{-3}{\bf q}\cdot({\bf p'}\times {\bf p_e}) ~, \\[2mm]
\delta f_+ &\eq M_K^2(q W)^{-1}\left[f_+(t)-f_+(W^2)\right] ~, \\[2mm]
N &= {16 \pi\alpha G_F^2|V_{us}|^2M_K^2} ~.
\end{split} \eeq
With this convention for $N$, the right hand
side in \eqref{eq:traces} is dimensionless. 
In the limit $m_e=0$, we immediately have
\beq
b_3\eq b_3^5\eq c_3\eq c_3^5\eq d_3\eq d_3^5 \eq e_3\eq e_3^5 \eq 0~.
\eeq
We use the abbreviations
\beq
\begin{array}{lllll}
z\, p   p'   =a \scs&  
z\, p   q    =b \scs&
z\, p   p_e  =c \scs&  
z\, p   p_\nu=d \scs\\[2mm]  
z\, p'  q    =e \scs&  
z\, p'  p_e  =f \scs&  
z\, p'  p_\nu=g \scs&
z\, q   p_e  =h \scs\\[2mm] 
z\, q   p_\nu=j \scs&  
z\, p_e p_\nu=k \scs& 
z\, p   W    =l \scs&  
z\, p'  W    =m \scs\\[2mm]
z\, q   W    =n \scs& 
z\, M_\pi^2=r   \scs& 
z\,=M_K^{-2}    \scs&
\end{array}
\eeq
and decompose all the coefficients according to 
$a_i = \hat a_i \, \bar a_i$ etc., where the prefactors 
$\hat a_i\,,\;\hat b_i\,\ldots$ are collected in
Table~\ref{tab:factors}. 
\begin{table}
\centering
\caption{Prefactors that  multiply the $\bar a_i\,,\;\bar b_i$ etc.
  \label{tab:factors}
}
\medskip
\renewcommand{\arraystretch}{1.4}
\begin{tabular}{cccccc}
\hline
$\hat a_1$ & $4/(e^2\,h)$ &
$\hat b_1$ & $4/(e\,h)$ & $\hat b^5_1$ & $4/h$      \\
$\hat a_2$ & $1/(e^2\,h)$ &
$\hat b_2$ & $4/(e\,h)$ & $\hat b^5_2$ & $4/(e\,h)$ \\
$\hat a_3$ & $4/e^2$      &
$\hat b_4$ & $1/(e\,h)$ & $\hat b^5_4$ & $2/h$      \\
\hline
$\hat c_1$ & $4/e$      & $\hat c^5_1$ & $1$        &
$\hat d_1$   & $4/(e\,h)$ \\
$\hat c_2$ & $1/e$      & $\hat c^5_2$ & $4\,n/e$   &
$\hat d_2$   & $4/(e\,h)$ \\
$\hat c_4$ & $1/e$      & $\hat c^5_4$ & $1$        &
$\hat d_4$   & $2/(e\,h)$ \\
\hline
$\hat d^5_1$ & $1/h$      & $\hat e_1$ & $4/e$    & $\hat e^5_1$ & $1$      \\
$\hat d^5_2$ & $4/(e\,h)$ & $\hat e_2$ & $4/e$    & $\hat e^5_2$ & $4\,n/e$ \\
$\hat d^5_4$ & $2/h$      & $\hat e_4$ & $4\,n/e$ & $\hat e^5_4$ & $1$      \\
\hline
\end{tabular}
\renewcommand{\arraystretch}{1.0}
\end{table}
We obtain the following expressions for the coefficients $\bar a_i$,
$\bar b_i$ and so on:
\bea
\bar a_1 &=& 2\,b\,d\,e\,(e + f) - e\,\bigl[h\,(2\,a\,d - g) +j\,(e + f) 
\nnnl && 
+ k\,(e + 2\,f)\bigr] 
+ h\,k\,r + 2\,c\,d\,(e^2 + 2\,e\,f - h\,r)\scs
\nnnl[0.75mm]
\bar a_2 &=&4\,e\,\bigl\{e\,k\,\bigl[2\,d\,(2\,c-h)+h-j\,(1+2\,c)-2\,k
\nnnl && 
+ 2\,b\,(d - h + k)\bigr] 
-  2\,e\,h\,l\,(d - j) 
\nnnl && 
+ 2\,h\,m\,\bigl[d\,h + c\,(j-2\,d) + k\,(1 - b)\bigr]\bigr\} 
\nnnl &&
- 4\,n\,\bigl\{2\,b\,d\,e\,f + e\,h\,\bigl[2\,d\,(-a + e + f) +g\bigr]  
\nnnl &&
- e\,f\,j - 2\,e\,k\,(f + a\,h) +  2\,h\,k\,r 
\nnnl &&
+ 2\,c\,(2\,d\,e\,f + e\,g\,h - 2\,d\,h\,r)\bigr\}\scs
\nnnl[0.75mm]
\bar a_3 &=& -2\,e^2\,k\,\bigl[2\,(c - h)\,(d - j) + k\,(2\,b-1)\bigr] 
\nnnl && 
- 2\,e\,n\,\bigl\{e\,k\,(c + d - h - j - l)
\nnnl &&
+ m\,\bigl[d\,h + c\,(j-2\,d) 
+ k\,(1 - b)\bigr]\bigr\} 
\nnnl && 
+ n^2\,\bigl[2\,e\,(d\,f + c\,g - a\,k) + r\,(k-2\,c\,d)\bigr]\scs
\nnnl[0.75mm]
\bar b_1 &=&e\,\bigl[d\,h\,(e + 2\,f) + g\,h\,(b+c)+j\,(c\,f-a\,h)
\nnnl && 
-k\,(b\,f+a\,h)\bigr] 
- h\,r\,(d\,h + c\,j - b\,k)\scs
\nnnl[0.75mm]
\bar b_2&=& j\,(c\,e\,k - e\,h\,l - c\,h\,m) + b\,k\,\bigl[e\,(h - k) + h\,m\bigr] 
\nnnl && 
+ n\,h\,(c\,g - a\,k)  
+ d\,h\,\bigl[e\,k - h\,m + n\,(e + f)\bigr]\scs
\nnnl[0.75mm]
\bar b_4&=& 2\,e\,\bigl\{c\,e\,k\,(e - 2\,g) + e\,\bigl[-b\,k\,(f + g) 
\nnnl && 
+ a\,k\,(j-h + 2\,k) 
+ l\,(g\,h + f\,j - e\,k)\bigr]
\nnnl &&  
+ c\,m\,(2\,g\,h - e\,j) + k\,m\,(b\,e - 2\,a\,h) 
\nnnl &&
+ d\,(e + 2\,f)\,(-e\,k + h\,m)\bigr\} 
\nnnl &&  
+ 2\,n\,\bigl\{-\bigl[e\,(c\,g\,(e - 2\,f) 
- 2\,b\,f\,g 
\nnnl &&  
+ a\,(-e\,k + 2\,f\,(j + k)))\bigr] 
+ r\,\bigl[c\,(-2\,g\,h + e\,j)
\nnnl &&
+k\,(- b\,e + 2\,a\,h)\bigr] 
+ d\,(e + 2\,f)\,(e\,f - h\,r)\bigr\}\scs
\nnnl[0.75mm]
\bar b^5_1&=& h\,(d\,e + b\,g + c\,g)-j\,(c\,f + a\,h)+k\,( b\,f - 
   a\,h)\scs
\nnnl[0.75mm]
\bar b^5_2&=& c\,e\,k\,(j-h) - b\,k\,\bigl[h\,(e + f - g) + e\,k\bigr] 
\nnnl &&  
+ h\,\bigl[a\,k\,(h - j) 
+l\,( - g\,h + e\,j + f\,j + e\,k) -
 d\,e\,n\bigr]\scs
\nnnl[0.75mm]
\bar b^5_4&=&
k\,\bigl\{-d\,e^2+a\,(e-2\,f)\,(j-h)
\nnnl &&  
+b\,\bigl[2\,f\,(g-f)-e\,(g+f)\bigr]\bigr\} 
+k^2\,( - 2\,r\,b + 2\,a\,e)
\nnnl &&  
+l\,\bigl[g\,h\,(e-2\,f)+(e+2\,f)\,(f\,j-e\,k) 
+ 2\,r\,h\,k\bigr]
\nnnl &&
+ m\,(d\,e\,h + b\,e\,k + 2\,b\,f\,k - 
   2\,a\,h\,k) 
\nnnl &&  
+ n\,\bigl[f\,(d\,e + 2\,b\,g - 2\,a\,j) 
-r\,(b\,k + d\,h)+ a\,e\,k\bigr] 
\nnnl &&  
+ c\,\bigl\{k\,(e^2 - 2\,r\,h + 2\,r\,j) 
+ m\,(2\,g\,h - 2\,f\,j) 
\nnnl &&  
+ r\,j\,n  
+ e\,\bigl[2\,f\,k - j\,m - g\,(2\,k + n)\bigr]\bigr\}\scs
\nnnl[0.75mm]
\bar c_1&=& 
e\,\bigl\{e\,k\,(d - h - j - l) +m\, (2\,h\,j + b\,k-d\,h) 
\nnnl &&  
+ c\,(e\,k - j\,m)\bigr\} 
- e\,n\,\bigl[g\,(c + h) 
+ f\,(d + j) - a\,k\bigr] 
\nnnl &&
+ n\,r\,(d\,h + c\,j - b\,k)\scs
\nnnl[0.75mm]
\bar c_2&=&8\,e\,k\,\bigl[2\,h\,j + b\,k - d\,h - c\,j\bigr] 
\nnnl &&
+  4\,n\,\bigl\{e\,k\,(c + d - 2\,(h + j) - l) 
\nnnl &&  
+ m\,(d\,h + c\,j - b\,k)\bigr\} - 4\,n^2\,(d\,f + c\,g - a\,k)\scs
\nnnl[0.75mm]
\bar c_4&=&8\,e^2\,k\,(d\,f + c\,g - g\,h - f\,j - a\,k + e\,k) 
\nnnl && 
+  4\,e\,n\,\bigl[e\,k\,(f + g - 2\,m)  
\nnnl &&
+m\, (-2\,d\,f - 2\,c\,g + g\,h + f\,j + 2\,a\,k)\bigr] 
\nnnl && 
+  4\,n^2\,\bigl[r\,(d\,f + c\,g +k\, (e-a)) -2\,e\,f\,g\bigr]\scs
\nnnl[0.75mm]
\bar c^5_1&=&4\,\bigl[b\,k\,(g-f) - e\,h\,k + a\,k\,(h - j) + e\,j\,k 
\nnnl &&  - g\,h\,l 
+  f\,j\,l - g\,h\,n + f\,j\,n\bigr]\scs
\nnnl[0.75mm]
\bar c^5_2&=& b\,k\,(f - g) - k\,(a - 2\,e)\,(h - j) +l\, (g\,h -f\,j)\scs
\nnnl[0.75mm]
\bar c^5_4&=& 4\,n\,\bigl[e\,k\,(g-f) + r\,k\,(h - j) +m\,(f\,j- g\,h)\bigr]
\fs \nnnl
\eea
\noindent The coefficients for the $T$-odd terms are
\beq
\begin{array}{ll}
\bar d_1=- e\,f + r\,h ~,& ~
\bar d_2= h\,(f + g + h + j) + e\,(h - k) ~,\\[1.8mm]
\mcl{
\bar d_4= 2\,e^2\,g + 2\,e\,h\,(f + g) - r\,e\,(h + j) - 2\,r\,h\,(h+j)
~,}
\\[1.8mm]
\bar d^5_1=4\,f ~,& ~
\bar d^5_2= h\,(j-f + g - h) - e\,(h + k) ~,\\[1.8mm]
\mcl{
\bar d^5_4=g\,(2\,e + 4\,f)+2\,h\,( g-f )- r\,(h + j + 2\,k)
~,}
\end{array} 
\eeq
and
\beq
\begin{array}{ll}
\bar e_1=  - r\,n + e\,(f + g) ~,&~
\bar e_2=2\,e\,k -n\,( f + g + h + j)~,\\[1.8mm]
\mcl{
\bar e_4=r\,(h + j) - e\,(f +g)~,
}
\\[1.8mm]
\bar e^5_1=4\,(g - f)~,&~
\bar e^5_2= f - g + h - j~,\\[1.8mm]
\mcl{
\bar e^5_4=4\,n\,(f - g) ~.
}
\end{array}
\eeq

\subsection{On the relation between \boldmath{$K_L$} and \boldmath{$K^0$} decays}

Here we comment on  the decay $K_L\rightarrow \pi^\pm e^\mp \nu_e\gamma$
and its relation to $K^0\rightarrow \pi^- e^+ \nu_e\gamma$ in light of the
contributions proportional to $\xi$ in \eqref{eq:traces}.
We neglect ${CP}$-violating contributions and write
\beq
|K_L\rangle \eq \frac{1}{\sqrt{2}}\left(|K^0\rangle - |\bar K^0\rangle\right)\fs
\eeq
The width for $K_L\rightarrow \pi^\pm e^\mp \nu_e\gamma$ 
is proportional to 
\beq
\int d_{\rm LIPS}\bigl(C_1+C_2\bigr) ~,
\eeq
where
\beq \begin{split}
C_1 &\eq \sum_{\rm spins} \bigl|T\bigl(K_L \to \pi^- e^+     \nu_e \gamma\bigr)\bigr|^2 ~,\\
C_2 &\eq \sum_{\rm spins} \bigl|T\bigl(K_L \to \pi^+ e^- \bar\nu_e \gamma\bigr)\bigr|^2 ~.
\end{split} \eeq
 In $C_1\, (C_2)$, only the component $|K^0\rangle$ ($|\bar
K^0\rangle$) contributes. We use $CP$ to transform the second term to the
first one. Doing so, 
all three-momenta of the particles change sign. Therefore, terms proportional
to $\xi$  drop out in the sum $C_1+C_2\,$, and
the decay width for $K_L\rightarrow \pi^\pm e^\mp \nu_e\gamma$
agrees with the one for 
$K^0\rightarrow e^+\nu_e\pi^-\gamma\,$, because in this decay, $\xi$ 
drops out as well after integration over the  momenta.
 These remarks remain true in the presence of the kinematical 
cuts considered in the main text, in connection with the ratio $\rR$.
 Therefore, up to terms quadratic in the structure-dependent 
terms, only the real parts of the amplitudes $V_i,A_i$ 
occur in the width and in $\rR$. Finally, the decay width 
for $K^0\rightarrow \pi^-e^+\nu_e$ coincides with $\Gamma(K_{e3})$. 
This leads to the expression \eqref{r} for the ratio $\rr$.

For $T$-odd terms in the context of $K^+_{e3\gamma}$ decays, see
\cite{braguta}.

\setcounter{figure}{0}
\setcounter{table}{0}
\section[Invariant amplitudes for $\kl3g^0$ at order $p^4$]
{Invariant amplitudes for \boldmath{$\kl3g^0$} at order \boldmath{$p^4$}}
\label{app:Vi}

In this appendix, we wish to give a simplified form of the $\kl3g^0$
one-loop amplitudes that is nevertheless as accurate as the exact result 
(that can be found in~\cite{BEG}) for all practical purposes.  
As the structure-dependent terms start to contribute at $\Order(q)$, we intend
to retain \emph{only} terms of order linear in the photon momentum and neglect
everything that is $\Order(q^2)$ or higher.  In this approximation, all the
structure functions $V_{1/2/3}$ can be written in terms of the
conventional two-point function $\bar{J}(t)$ plus chiral logarithms and rational
functions of the masses.  
As remarked before, $V_{4}=0$ at this order.
We use the following definitions and conventions:
\begin{align}
M_1 &\eq M_K    ~,~~ m_1 \eq M_\pi ~,~~
M_2  \eq M_\eta ~,~~ m_2 \eq M_K   ~, \no\\
\Sigma_i &\eq M_i^2+m_i^2 ~,~~
\Delta_i  \eq M_i^2-m_i^2 ~,
\end{align}
the K{\"a}ll{\'e}n function
\begin{align}
\lambda_i(t) &\eq \lambda(t,M_i^2,m_i^2) \\ &\eq
t^2+M_i^4+m_i^4-2\left(t\left(M_i^2+m_i^2\right)+M_i^2 m_i^2\right)
~, \no
\end{align}
and the loop functions
\begin{align}
\bar{J}_1(t) &\eq \bar{J}_{K\pi}(t) ~,\quad
\bar{J}_2(t) \eq \bar{J}_{\eta K}(t) ~, \no\\[2mm]
\bar{J}_{ab}(t) &\eq J_{ab}(t) - J_{ab}(0) ~,\label{defJbar}\\
J_{ab}(q^2) &\eq 
\frac{1}{i}\int \frac{d^dl}{(2\pi)^d}  
  \frac{1}{\bigl(M_a^2-l^2\bigr)\bigl(M_b^2-(l-q)^2\bigr)} ~. \no
\end{align} 
Our results can be written as follows:
\bea
V_1 &=& - \frac{8}{F^2}\bar{L}_9 
\\ &-& 
\frac{1}{4F^2t} \sum_{i=1}^2 \biggl\{
\biggl(\frac{2\lambda_i(t)}{t}+3\Sigma_i+\left(M_K^2-M_\pi^2\right)\biggr)\bar{J}_i(t)
\no\\ && 
+ \frac{t\Sigma_i-2M_i^2m_i^2}{16\pi^2\Delta_i}\log\frac{m_i^2}{M_i^2}
+ \frac{t-3\Sigma_i}{48\pi^2} \biggl\} + \Order(q) ~,
\no\\
V_2 &=& \frac{4}{F^2} \left(\bar{L}_9 + \bar{L}_{10}\right) 
\\ &-&
\frac{1}{2F^2} \biggl\{ \frac{M_K^2+M_\pi^2}{t}\bar{J}_{1}(t) 
-\frac{1}{16\pi^2} 
+\frac{4M_K^2M_\pi^2}{\lambda_1(t)} \times
\no\\ && \quad
 \times \biggl(\bar{J}_1(t)-\frac{1}{16\pi^2}
         -\frac{t-\left(M_K^2+M_\pi^2\right)}{32\pi^2\left(M_K^2-M_\pi^2\right)}
          \log\frac{M_K^2}{M_\pi^2} \biggr) \biggr\} 
\no\\&+&
\frac{1}{2F^2t} \sum_{i=1}^2 \biggl\{
\frac{2M_i^2m_i^2}{\lambda_i(t)}
  \Bigl(2t+3\left(M_K^2-M_\pi^2\right)+\Delta_i\Bigr) \times
\no\\ && \qquad \times
  \Bigl(\bar{J}_i(t)-\frac{1}{16\pi^2}\Bigr) 
\no\\ && +
\biggl(3m_i^2+\frac{2m_i^2\Delta_i}{t}+\frac{3(M_K^2-M_\pi^2)\Delta_i^2}{t^2}\biggr)
 \bar{J}_i(t) 
\no\\ &&
+\frac{M_i^2m_i^2}{16\pi^2\Delta_i} 
 \log\frac{m_i^2}{M_i^2} \biggl[
 3\left(1-\frac{M_K^2-M_\pi^2}{t}\right) +\frac{2t\Sigma_i}{\lambda_i(t)} 
\no\\ && 
+ \frac{3(M_K^2-M_\pi^2)+\Delta_i}{2\lambda_i(t)} \times
\no\\ && \qquad \times
\left(2(t-\Sigma_i)-\Delta_i+\frac{3}{10}\left(M_K^2-M_\pi^2-\Delta_i\right)\right)
\biggr] 
\no\\ &&
-\frac{1}{16\pi^2}\biggl(t-m_i^2+\frac{M_K^2-M_\pi^2}{2t}
  \bigl(3\Sigma_i+t\bigr)\biggr)
 \biggr\} + \Order(q) ~,
\no\\
V_3 &=& \frac{1}{2F^2t} \sum_{i=1}^2 \biggl\{
  \frac{1}{t^2}\Bigl(\left(M_K^2-M_\pi^2\right)t+6\Delta_i^2\Bigr) \bar{J}_i(t)
\\ &&
- \frac{\bigl(3\Sigma_i+(M_K^2-M_\pi^2)\bigr)(t-\Sigma_i)-12M_i^2m_i^2}{2\lambda_i(t)}
\times
\no\\ && \qquad \times
\Bigl( \bar{J}_i(t)-\frac{1}{16\pi^2}\Bigr) 
\no\\ &&
- \frac{M_i^2m_i^2}{16\pi^2t\Delta_i} \biggl(
\frac{(M_K^2-M_\pi^2)t+3\Delta_i^2}{\lambda_i(t)}+3\biggr) 
   \log\frac{m_i^2}{M_i^2}
\no\\ &&
+ \frac{t-6\Sigma_i}{32\pi^2t} \biggr\} 
+ \Order(q) ~. \no
\eea
Note that this expansion necessarily upsets the analytic structure as the cuts
in the variables $t$ and $W^2$ coincide in the limit of vanishing photon
momentum.  However, these cuts lie far outside the physical region (see
discussion in Sect.~\ref{sec:p6}).  Furthermore, despite their appearance,
all the functions above are regular and smooth at $t=0$ and $t=(M_K-M_\pi)^2$.
The results of the even further simplification by setting $t=0$ are displayed
in the main text \eqref{Vit=0}.

\setcounter{figure}{0}
\setcounter{table}{0}
\section[Axial form factors at order $p^6$]
{Axial form factors at order \boldmath{$p^6$}}\label{app:axial}

\begin{figure}
\vskip 2mm
\centering
\includegraphics[width=8.5cm]{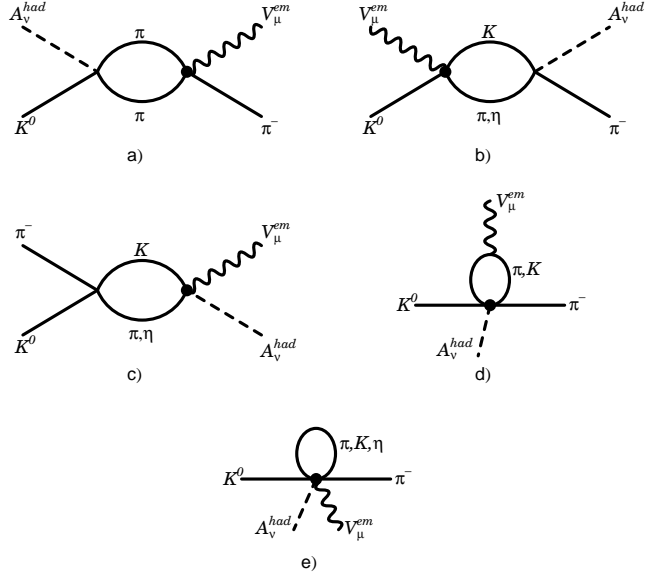}
\caption{Diagrams that contribute to the anomalous amplitude
  $A_{\mu\nu}$ at order $p^6$ [in a), the contribution from $\pi,\eta$
  intermediate states vanishes]. 
  Charges of the mesons running in the loops  are not indicated.  
  The filled vertices denote a contribution from the anomalous
  Lagrangian at order $p^4$. External line insertions in the tree
  diagram of order $p^4$ are not displayed.\label{fig:axialloops}}
\end{figure}
In this appendix, we give the explicit formulae for the next-to-leading order
corrections to the axial form factors $A_1$, $A_2$, and $A_4$, as 
written out formally in \eqref{A1struct}-\eqref{A4struct}.
The necessary loop diagrams for this calculation are displayed in
Fig.~\ref{fig:axialloops}.
We find the following combinations of loop functions and
counterterms:
\begin{align}
S_1(s) &\eq 2 H^r_{\pi\pi}(s) + \frac{16\pi^2}{3}\, C^{r}_{1s} \,s~,
\label{eq:axialstu:beg}\\
S_2(s) &\eq \frac{16\pi^2}{3} \,C_{2s} \,s~,\\
T_1(t) &\eq   H^r_{K  \pi}(t) +  H^r_{\eta K}(t) +\frac{16\pi^2}{3} \,C^{r}_{1t} \,t~,\\
T_2(t) &\eq T_{K  \pi}^r(t)   + T_{\eta K}^r(t) +\frac{16\pi^2}{3} \,C^{r}_{2t} \,t ~,\\
U_1(u) &\eq -  H^r_{K\pi }(u) -3 H^r_{\eta K}(u) +\frac{16\pi^2}{3} \,C^{r}_{1u} \,u~,\\
U_2(u) &\eq 2 H^r_{K  \pi}(u) +6 H^r_{\eta K}(u) +\frac{16\pi^2}{3} \,C^{r}_{2u} \,u~,\\
X_1    &\eq            2\mu_\pi -\mu_K -\mu_\eta 
  + \frac{16\pi^2}{3}\bigl(C^{r}_{1\pi} M_\pi^2 + C^{r}_{1K} M_K^2  \bigr)~, \\
X_2    &\eq -\frac{31}{12}\mu_\pi +\frac{19}{6}\mu_K +\frac{3}{4}\mu_\eta  \no\\
  & \hskip 0.5cm
  + \frac{16\pi^2}{3}\bigl(C^{r}_{2\pi} M_\pi^2 + C^{r}_{2K} M_K^2  \bigr)~. ~~
\label{eq:axialstu:end}
\end{align}
The loop function $H^r_{ab}(x)$ is defined as
\beq \begin{split}
H^r_{ab}(x) &=
\frac{1}{12F^2}\biggl\{ \frac{\lambda(x,M_a^2,M_b^2)}{x}\bar{J}_{ab}(x) 
 + \frac{x-3\Sigma_{ab}}{24\pi^2} 
\\ & 
 - \frac{x}{32\pi^2}\log\frac{M_a^2M_b^2}{\mu^4} 
 - \frac{x\,\Sigma_{ab}-8M_a^2M_b^2}{32\pi^2\Delta_{ab}}\log\frac{M_a^2}{M_b^2} 
\biggr\} ~. \label{Hdef} 
\end{split} \eeq
The other functions can also be written in relatively compact forms:
\begin{align}
T_{K\pi}^r(t) &\eq \frac{1}{24F^2} \biggl\{ 
13t \biggl[ \bar{J}_{K\pi}(t) 
\no\\ & \hskip 1.0cm 
-\frac{1}{32\pi^2}\Bigl(\log\frac{\mk\mpi}{\mu^4}
+ \frac{\Sigma_{K\pi}}{\Delta_{K\pi}} \log\frac{\mk}{\mpi} \Bigl) \biggr]
\no\\ & \hskip 0.5cm
-
\biggl[
2\Sigma_{K\pi}+16\Delta_{K\pi} 
\no\\ & \hskip 1cm
-\Bigl(8\Sigma_{K\pi}-11\Delta_{K\pi}
  +\frac{8\Delta_{K\pi}^2}{t}\Bigr)\frac{\Delta_{K\pi}}{t} \biggr] 
\bar{J}_{K\pi}(t) 
\no\\ & \hskip 0.5cm
+\frac{M_K^2M_\pi^2(2\Delta_{K\pi}+t)}{4\pi^2t\Delta_{K\pi}}\log\frac{\mk}{\mpi}
\no\\ & \hskip 0.5cm
- \frac{(t-3\Sigma_{K\pi})(t-\Delta_{K\pi})}{12\pi^2t} \biggr \}~,  \\
T_{\eta K}^r(t)  &\eq \frac{1}{24F^2} \biggl\{ 
t \biggl[ \bar{J}_{\eta K}(t) 
\no\\ & \hskip 1cm
-\frac{1}{32\pi^2}\Bigl(\log\frac{\me\mk}{\mu^4}
+ \frac{\Sigma_{\eta K}}{\Delta_{\eta K}} \log\frac{\me}{\mk} \Bigl) \biggr]
\no\\ & \hskip 0.5cm
+\biggl[
2\Sigma_{\eta K}+8\Delta_{\eta K}
\no\\ & \hskip 1cm
-\Bigl(\frac{8}{3}\Sigma_{\eta K}+9\Delta_{\eta K}-\frac{8\Delta_{\eta K}^2}{t}\Bigr)\frac{\Delta_{K\pi}}{t}
 \biggr] 
\bar{J}_{\eta K}(t) 
\no\\ & \hskip 0.5cm
+\frac{\me\mk(2\Delta_{K\pi}-t)}{4\pi^2t\Delta_{\eta K}}\log\frac{\me}{\mk}
\no\\ & \hskip 0.5cm
- \frac{(t-3\Sigma_{\eta K})(t-\Delta_{K\pi})}{12\pi^2t} \biggr \}~,  \\
\mu_a &\eq \frac{M_a^2}{32\pi^2F^2}\log\frac{M_a^2}{\mu^2} ~,
\end{align}
where we have used $\Sigma_{ab}=M_a^2+M_b^2$, $\Delta_{ab}=M_a^2-M_b^2$,
and the two point function $\bar{J}_{ab}(x)$ as defined in
\eqref{defJbar}.
The combinations of low-energy constants occurring in
\eqref{A1struct}--\eqref{A4struct} 
and \eqref{eq:axialstu:beg}--\eqref{eq:axialstu:end} 
are given in Table~\ref{tab:coeff_cis}
according to the numbering in~\cite{p6anom}.
\begin{table}
\caption{The coefficients from
  \eqref{eq:axialstu:beg}--\eqref{eq:axialstu:end}  
  in terms of the
  renormalized low-energy constants $C_{i}^{Wr}$. For example,
  $C_{1s}^r=4C_{13}^{Wr}-10 C_{14}^{Wr}+\ldots$. 
  Constants without superscript $r$ are scale independent.
  \label{tab:coeff_cis}
}
\small{
\bea
{
\renewcommand{\arraystretch}{1.4}
\begin{array}{lrrrrrrrrrrr}
\hline
&
C_{1s}^r   &
C_{1t}^r   &
C_{1u}^r   &
C_{1\pi}^r &
C_{1K}^r   &
C_{2s}     &
C_{2t}^r   &
C_{2u}^r   &
C_{2\pi}^r &
C_{2K}^r   &
C_{4A} \\
\hline
C_{2}^{Wr}  & 0 & 0 & 0 & 24&-24& 0 & 0 & 0 &-48& 48& 0
\\
C_{4}^{Wr}  & 0 & 0 & 0 & 8 &-16& 0 & 0 & 0 &-16& 0 & 0
\\
C_{5}^{Wr}  & 0 & 0 & 0 &-8 & 16& 0 & 0 & 0 & 16& 0 & 0
\\
C_{7}^{W}   & 0 & 0 & 0 & 0 & 0 & 0 & 0 & 0 &-16&-32& 0
\\
C_{9}^{W}   & 0 & 0 & 0 & 0 & 0 & 0 & 0 & 0 & 0 & 48& 0
\\
C_{11}^{Wr} & 0 & 0 & 0 & 0 & 0 & 0 & 0 & 0 & 0 &-48& 0
\\
C_{13}^{Wr} & 4 & 4 & 4 &-10& 2 &-8 &-10& 0 & 22& 10&-16
\\
C_{14}^{Wr} &-10&-10&-4 & 12& 18& 20& 16& 0 &-20&-16& 48
\\
C_{15}^{Wr} & 8 & 8 & 8 &-12&-12&-16&-20& 0 & 28& 20&-32
\\
C_{19}^{Wr} & 2 & 2 & 2 &-2 &-2 &-4 &-2 & 0 & 2 & 2 & 0
\\
C_{20}^{Wr} &-2 &-2 &-8 & 8 & 2 & 4 & 8 & 0 &-20&-8 & 16
\\
C_{21}^{Wr} & 4 & 4 & 4 &-4 &-4 &-8 &-4 & 0 & 4 & 4 & 0
\\
C_{22}^{Wr} &-1 &-1 &-4 & 4 & 1 & 6 & 4 & 8 &-10&-4 & 8
\\
C_{23}^{W}  & 9 & 9 & 6 &-6 &-9 &-18&-12&-12& 6 & 12& 8
\\ \hline
\end{array}
}
\renewcommand{\arraystretch}{1.0}
\nonumber
\eea
}
\end{table}

\setcounter{figure}{0}
\setcounter{table}{0}
\section[Inner bremsstrahlung in $\kl3g^0$ decays]
{Inner bremsstrahlung in \boldmath{$\kl3g^0$} decays}\label{app:WI}

In this appendix we discuss the separation of the hadronic tensor
$V_{\mu\nu}$ into an IB and a SD part. To be specific, we  imagine a
calculation in the framework of ChPT to all orders and discuss the
decomposition there.
\begin{figure}[hb]
  \centering
  \includegraphics[width=7.5cm]{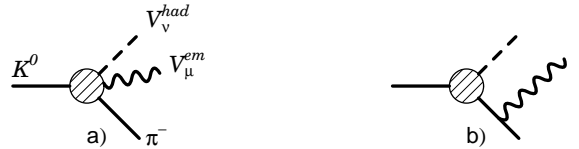}
  \caption{Diagrams for $\vmn$, evaluated in the framework of ChPT. The
    hatched blobs denote one-particle irreducible graphs.}
  \label{fig:IB}
\end{figure}
 The relevant diagrams can be grouped in two classes,  displayed in
 Fig.~\ref{fig:IB}. The hatched blobs denote one-particle irreducible
 contributions. The
 diagram b) generates a pole in 
the variable $s=(p'+q)^2$, at $s=M_\pi^2$, corresponding to the
intermediate pion  state.
We isolate the contribution of this pole by
writing 
\beq
V_{\mu\nu} \eq \tilde{V}_{\mu\nu} + \frac{p'_\mu}{p'q}\,
\left[2p_\nu f_+(W^2) - W_\nu f_2(W^2) \right] ~,
\label{pole}
\eeq
where $W=p-p'-q$. 
 In the following, we assume that 
this is the only
singular part at $q=0$ in the tensor $V_{\mu\nu}$, or, in other words, that
$\tilde V_{\mu\nu}$ is regular at $q=0$. This is the only assumption in the
derivation of the final expression for the IB term. We have checked that it
is true at one-loop   order in ChPT, see below,
 and we  see no reason why it should
 not be correct to any order, and thus true in QCD.
Next, we write this regular part as
\beq \begin{split}
\tilde V_{\mu\nu} &\eq v_0\,g_{\mu\nu}+v_1\, p'_\mu q_\nu+
v_2\,W_\mu q_\nu  +v_3\, p'_\mu W_\nu  \\
& \hskip 0.45cm
+ v_4\, p'_\mu p'_\nu  + v_5\, W_\mu p'_\nu
 +v_6\, W_\mu W_\nu  ~. 
\end{split} \eeq
The Ward identity \eqref{WI} generates three
 conditions on $\tilde V_{\mu\nu}$,
\beq \begin{split}
v_0 + v_1\, p'q + v_2\, qW &\eq 2\triangle f_+-f_2 ~,\\
      v_3\, p'q + v_6\, qW &\eq 2\triangle f_+-\triangle f_2 ~,\\
      v_4\, p'q + v_5\, qW &\eq 2\triangle f_+ ~,
\end{split} \eeq
with
\beq
\triangle f_i \eq f_i(t)-f_i(W^2) ~.
\eeq
The first equation can be solved for $v_0$. Furthermore, we set
\beq
v_5 \eq \frac{2\triangle f_+}{qW}+\tilde v_5 ~,~~
v_6 \eq \frac{2\triangle f_+-\triangle f_2}{qW}+\tilde v_6 ~.
\eeq
and are  left with 
\beq \begin{split}
v_4\, p'q+\tilde v_5\, qW \eq 0 ~,\\
v_3\, p'q +\tilde v_6\,qW \eq 0 ~.
\end{split} \label{eq:twoequations} \eeq
At this stage, we use the fact that the Lorentz invariant amplitudes $v_i$ are
defined for any value of the kinematic variables $p'q,qW$, and that the
amplitudes are assumed to be non-singular at $p'q=0$.  It then follows
that $\tilde v_{5,6}$  are proportional to $p'q$,
\beq
\tilde v_{5,6} \eq -p'q \,\tilde v_{4,3} ~.
\eeq
where the sign and the numbering is  chosen for convenience. Finally, we obtain
\beq
v_{3,4} \eq qW \,\tilde v_{3,4} ~.
\eeq
Collecting the results, we find that $V_{\mu\nu}^{\rm SD}$ can be written in the form
displayed in \eqref{eq:ViSD}, with
\beq
(V_1,V_2,V_3,V_4) \eq (v_1,v_2,\tilde v_3,\tilde v_4) ~.
\eeq
Appendix~\ref{app:Vi} contains the explicit expression of
the form factors $V_i$ in the limit $q=0$, illustrating 
the fact that they indeed are 
non-singular at $q=0$ at next to leading order in
 ChPT, as mentioned above.

\end{appendix}


\end{document}